 \documentclass[final,authoryear,12pt,times]{elsarticle}

 \usepackage{graphicx}

\usepackage{amssymb}

\begin{document}

\begin{frontmatter}



\title{Study of open cluster King 13 using CCD VI, 2MASS and Gaia DR2 Astrometry.}


\author{Alok Durgapal\footnote{E-mail: alokdurgapal@gmail.com (Alok Durgapal); devendrabisht297@gmail.com (D. Bisht);
geetarangwal91@gmail.com (Geeta Rangwal); harmeenkaur.kaur229@gmail.com (Harmeen Kaur); rkant@aries.res.in (R. K. S Yadav)},
D Bisht\footnote{Corresponding author; Tel:+86-15695691278}, Geeta Rangwal$^1$, Harmeen Kaur$^1$, R. K. S. Yadav$^3$\\}

\address{$^{1}$Department of Physics, DSB Campus, Kumaun University, Nainital-263002, Uttarakhand, India\\
$^{2}$Key Laboratory for Researches in Galaxies and Cosmology, University of Science and Technology
             of China, Chinese Academy of Sciences, Hefei, Anhui 230026, China\\
$^{3}$Aryabhatta Research Institute of Observational Sciences, Manora Peak Nainital 263 002, India\\
}

\begin{abstract}

In this paper, we present astrophysical parameters of the open cluster King 13 based on the $VI$ CCD and 2MASS $JHK_{s}$
photometric data. This is a poorly studied cluster, for which new results have been found in the present work. To 
identify probable members, we use proper motion data from Gaia~DR2 catalogue. The mean proper motion of the cluster is
determined as $-2.8 \pm 0.2$ and $-0.88 \pm 0.14$ mas $yr^{-1}$ and cluster extent is derived as $3^{\prime}.2$. 
Using color-magnitude diagrams, we estimate the age and distance of the cluster as $510 \pm 60$ Myr and $3.84\pm 0.15$ kpc 
respectively. Interstellar reddening $E(B-V)$ in the direction of the cluster is determined as $0.80 \pm 0.2$ mag using
color-color diagram. Mass function slope of the cluster is found to be comparable with the Salpeter value. The total mass of
this cluster is derived as 270 $M_{\odot}$. The present analysis shows that King 13 is a dynamically relaxed cluster.

\end{abstract}


\end{frontmatter}



\section{Introduction}

Open clusters (OCs) are very helpful objects to understand the structure and stellar evolution of the Milky Way. Indeed,
each cluster includes stars of different mass, which were formed in the same cloud, i.e. they share the same age and chemical 
composition. The structure of most OCs can be roughly described by two subsystems, the dense core and the sparse halo \citep{2009MNRAS.397.1915B}. OCs have become fundamental probes of Galactic disk properties by their location on Galactic disk 
\citep{1982A&A...109..213L, 1994AJ....108.1773J, 1995ARA&A..33..381F, 2006A&A...445..545P, 2010AJ....140..954C}. The fundamental
parameters of an open cluster e.g. distance, age and interstellar extinction can be estimated by comparing color-magnitude diagram
(CMD) and color-color (CC) diagrams with the modern theoretical models. It is important to study the unexplored OCs to determine
their properties with the aim of improving the picture of the Galactic disk.

King 13 is positioned at $\alpha=00^{h} 10^{m} 06^{s}$ and $\delta=61^{\circ} 10^{\prime}$ (J2000.0), corresponding to Galactic
coordinates $l$= $117^{\circ}.96$ and $b$=$-1^{\circ}.3$. This object 
is located in the Perseus arm, in the second Galactic
quadrant of Milky Way. \citet{1979AN....300..295M} obtained UBV photographic photometry for King 13 and calculated its distance as 1.73 kpc.
\citet{2007MNRAS.377..829S} determined its distance as $3.10\pm 0.33$ kpc and its age as 300 Myr, while \citet{2011AcA....61..231B}
estimated cluster distance as $2.96\pm0.19$ kpc and age as 794 Myr. Therefore, there is a difference in parameters from one study to
another. So there is a need to re-visit this cluster and determine its parameters in a more accurate way, making use of new tools
and data. 

A new era in dynamical astronomy has begun with the second data release of the Gaia mission in 2018 April \citep{2016A&A...595A...1G}.
In the present analysis, we used CCD photometric data in $VI$ filters, 2MASS data in $JHK_{s}$ filters and Gaia DR2 proper motion data to
determine the fundamental and structural parameters of the cluster King 13.

The paper is organized as follows: Data reduction is presented in Section \ref{Obs}. Selection of cluster members is described in
Section \ref{prop}. Fundamental astrophysical parameters are estimated in Section \ref{ana}. Study of luminosity, mass function
and mass-segregation are given in Section \ref{mf}. Finally, we concluded our study in Section \ref{con}.

\section{Observational data and data reduction}
\label{Obs}

\begin{figure}
\begin{center}
\centering
\vspace{-1cm}
\includegraphics[width=8.5cm, height=8.5cm]{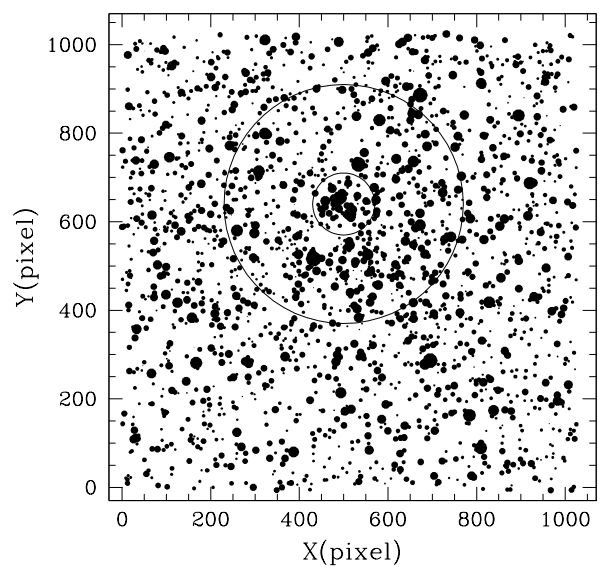}
\caption{Identification chart of the stars in the cluster and field regions of King 13. North is up and East in the right
direction. The outer circle represents the extent of the cluster
while the inner circle shows the extent of core of the cluster.
The smallest size denotes star of $V$ $\sim$ 20 mag.}
\label{id}
\end{center}
\end{figure}

The CCD broadband $VI$ images for King 13, were collected using a 2K$\times$2K CCD system at the f/13 Cassegrain focus 
of the 104 cm Sampurnanand telescope located at ARIES, Manora peak, Nainital, India. The CCD used for the present observations
has 24 $\mu$m square pixel size, resulting in a scale of 0$^{\prime\prime}$.36 pixel$^{-1}$ and a square field of view 
of 12$^{\prime}$.6 size. The CCD gain was 10 e$^{-}$/ADU while the readout noise was 5.3 e$^{-}$. Log of observations 
is listed in Table \ref{tab1}. Observations were taken in 2$\times$2 pixels binning mode to improve S/N ratio. The
identification map of the observed region for King 13 is shown in Fig. \ref{id}. In this figure, the core and cluster regions
are indicated by the inner and outer circle respectively.

Bias and twilight flats were also taken along with the target field. 
We have used IRAF software for the pre-processing of our observed
CCD images. In this step, we have done bias subtraction, flat field correction and removal of cosmic rays from science images. The subsequent
data reduction and analysis were done using the DAOPHOT software \citep{1987PASP...99..191S}. The stellar photometric routine of DAOPHOT
was used for the instrumental magnitude determination. Both PSF and aperture photometry was carried out to find the instrumental magnitudes
of the stars. The details of the processing of the images can be found in our previous papers \citep{1997A&AS..122..111P, 1997BASI...25..489D,
2001A&A...375..840D, 2016NewA...42...66B, 2019MNRAS.482.1471B}.

\subsection {Photometric calibration}

We observed the standard field SA 95 \citep{1922AJ....104..340L} in $V$ and $I$ filters for calibrating the observational data
of King 13. The 7 standard stars (SA95-41, 42, 43, 97, 102, 112, 115) used in the calibrations having brightness and
color range 12.77 $\le V \le$ 16.11 and $-0.329 < (V-I) < 1.448$ respectively, thus covering the bulk of the cluster stars. For
the extinction coefficients, we considered the values for the ARIES site 
\citep{2000BASI...28..675K}. The derived calibration equations
using least square linear regression for converting the instrumental magnitude into the standard magnitude, are as follows:\\

~~~~~~~~~~~~~~~~~~~~~~$v=V+Z_{V}+C_{V}(B-V)+k_{V}X$~~~~~~~~~~~~~~~(1)    \\

~~~~~~~~~~~~~~~~~~~~~$i=I+Z_{I}+C_{I}(V-I)+k_{I}X$~~~~~~~~~~~~~~~~(2)     \\

where $v$ and $i$ are the instrumental and $V$ and $I$ are the standard magnitudes, $X$ is the airmass. The color
coefficients $(C)$ and zero points $(Z)$ for different filters are listed in Table \ref{tab2}. The errors in zero
points using standard stars from Landolt SA 95 field and color coefficients are 0.01 mag. 

\begin{figure}
\begin{center}
\hbox{
\includegraphics[width=6.5cm, height=6.5cm]{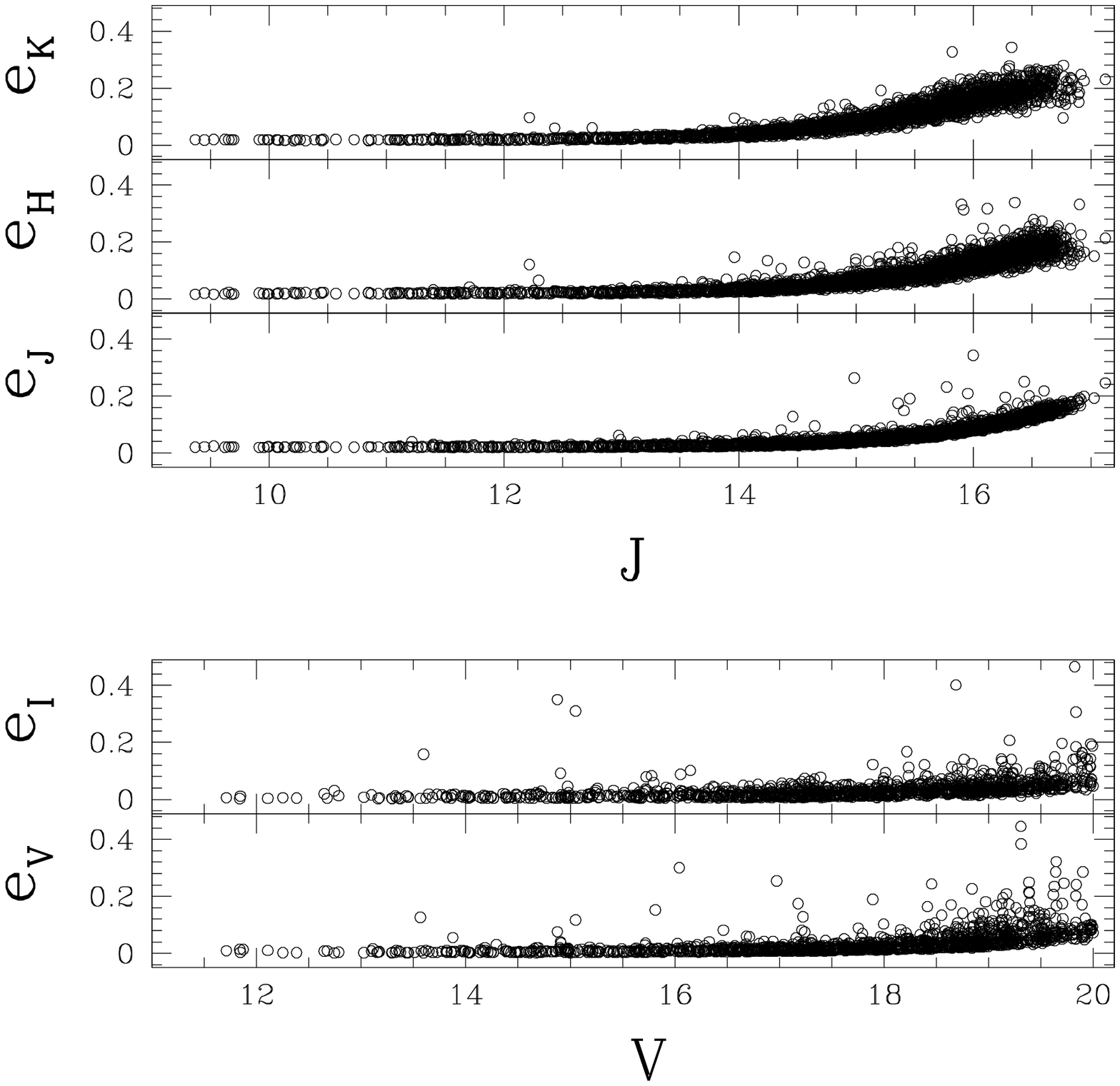}
\includegraphics[width=6.5cm, height=6.5cm]{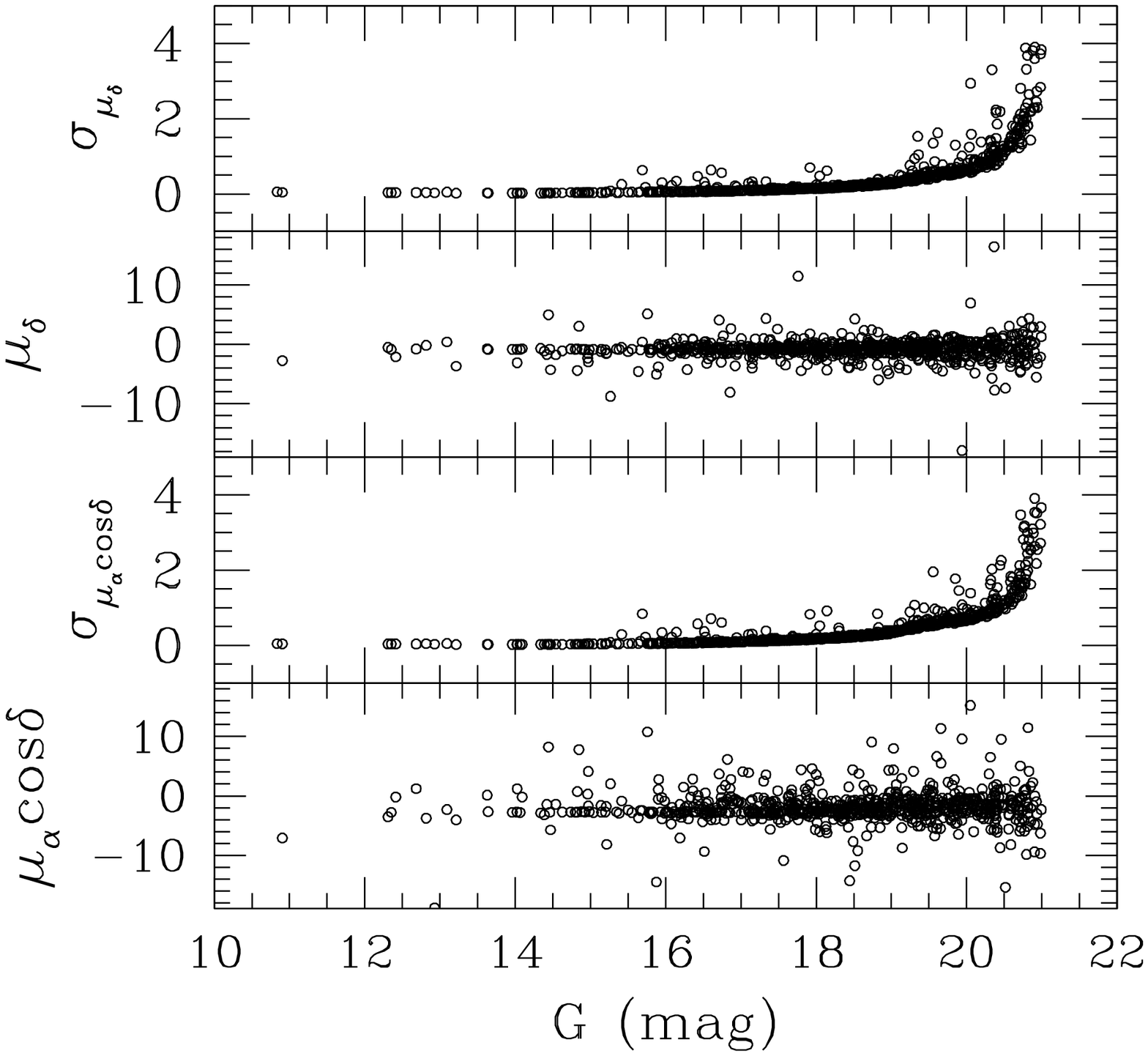}
}
\caption{Left panel shows photometric errors in $V$ and $I$ magnitudes
against $V$ magnitude and errors in $J$, $H$ and $K$ magnitudes against $J$
magnitudes. Right {\bf panel} shows the plot
of proper motions in both RA and DEC directions and their errors versus $G$ magnitude.}
\label{error}
\end{center}
\end{figure}

The internal errors derived from DAOPHOT are plotted against $V$ magnitude in Fig. \ref{error}. Photometric global
(DAOPHOT+Calibrations) errors are also estimated, which are listed in Table \ref{tab3}. For $V$ filter, the error 
is 0.05 mag at $V\sim$17 mag and 0.07 mag at $V\sim$20 mag.

\begin{figure*}
\begin{center}
\hbox{
\includegraphics[width=12.5cm, height=12.5cm]{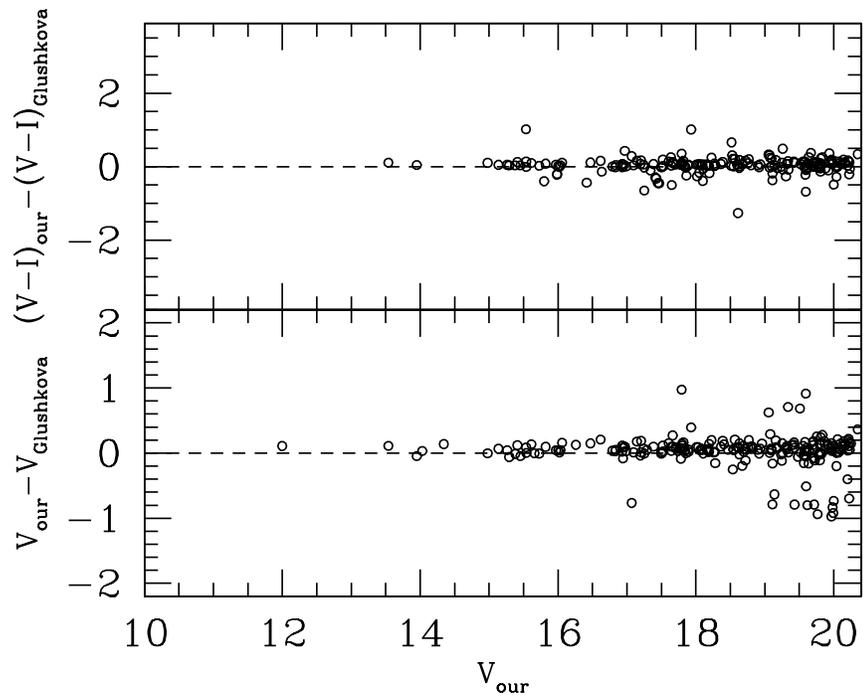}
}
\caption{A comparison of the present photometry with photometric
data of \citet{2010AstL...36...14G} for King 13. The open circles represent
difference of both the photometries as a function of present photometry.}
\label{comparision}
\end{center}
\end{figure*}

In Fig. \ref{comparision}, we present a comparison of present photometry with the photometric data of
\citet{2010AstL...36...14G}. In this figure, the difference between the two photometries is plotted as a function of
the present photometry. The dotted lines represent the zero difference between present photometry and photometry
by \citet{2010AstL...36...14G}. This shows that most of the data points lie near the zero difference line
so our photometric data is in good agreement with that of \citet{2010AstL...36...14G}. The mean difference and 
standard deviation per magnitude bin are also given in Table \ref{match_error}.

\subsection{Gaia DR2 data}

We used Gaia DR2 data \citep{2016A&A...595A...1G} to select the cluster members and determine to mean proper 
motion of cluster King 13. This data consist of five parametric astrometric solution, which includes positions on 
the sky $(\alpha, \delta)$, parallaxes and proper motion (PM) ($\mu_{\alpha} cos\delta, \mu{\delta}$) with a limiting
magnitude of $G \sim 21$ mag. Parallax uncertainties are in the range of up to 0.04 milliarcsecond (mas) for sources 
at $G\le15$ mag and $\sim$ 0.1 mas for sources at $G\sim17$ mag. The uncertainties in the respective proper motion 
components are up to 0.06 mas $yr^{-1}$ (for $G\le15$ mag), 0.2 mas $yr^{-1}$ (for $G\sim17$ mag) and 1.2 mas $yr^{-1}$
(for $G\sim20$ mag). The proper motion and their corresponding errors plotted against $G$ magnitude are shown in
Fig \ref{error}. This figure shows that errors in proper motion components are $\sim 1.2$ at $G\sim20$ mag.

\section{Mean proper motion and cluster membership}
\label{prop}

The contamination due to field stars always affects the determination of fundamental parameters of the cluster. Proper motion is one
of the tools to remove those field stars from the cluster main sequence. We used Gaia DR2 proper motion 
and parallax data to separate cluster members from non member stars.

\begin{figure}
\begin{center}
\includegraphics[width=10.5cm, height=10.5cm]{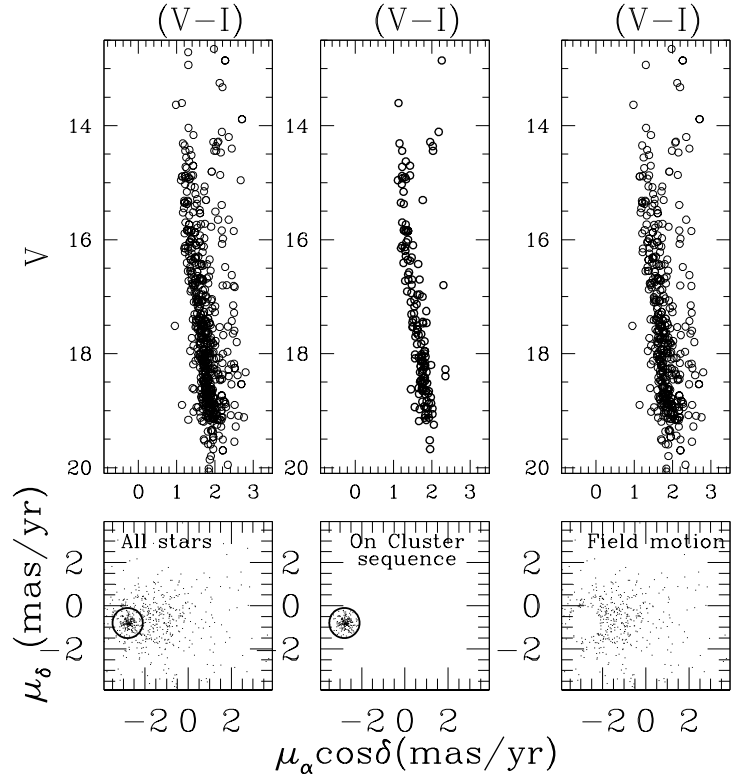}
\caption{Colour magnitude diagrams (CMDs) based on our $VI$ photometry (top panels) and proper
motion vector point diagrams (VPDs) based on {\it Gaia} DR2 data (bottom panels). Left panels display the
entire sample. Central panels display the candidate members (enclosed in a circle of radius $0.7~ mas~ yr^{-1}$
around the cluster center in VPD). Probable background/foreground filed stars in the direction of the cluster
are displayed in the right panels.}
\label{pm_dist}
\end{center}
\end{figure}

To see the distribution of member and non member stars, we plotted vector point diagram (VPD) in 
PMs $\mu_{\alpha} cos{\delta}$ and $\mu_{\delta}$ as shown in Fig \ref{pm_dist}. Corresponding $V$ versus
$(V-I)$ CMDs are also shown in top panels. The left panel shows all the observed stars, while the middle and 
right panel shows probable cluster members and field stars respectively.

\begin{figure}
\begin{center}
\includegraphics[width=10.5cm, height=10.5cm]{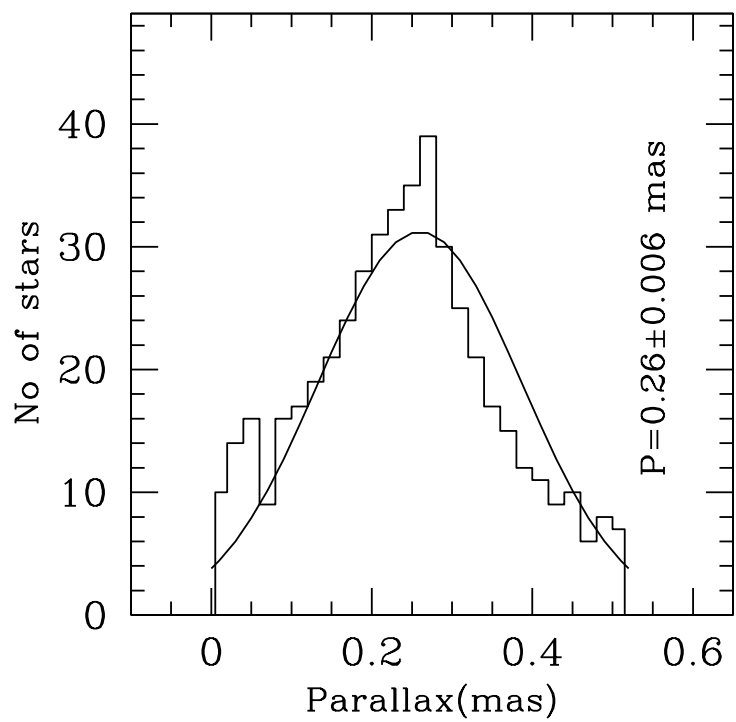}
\caption{The histograms for the estimation of mean parallax. The Gaussian function fit
to the central bins.}
\label{parallax}
\end{center}
\end{figure}

We determined the 
average value of parallax for stars inside the circle of
vector point diagram, by building a histogram of 0.15 mas bin as
shown in Fig \ref{parallax}. The average value of parallax is determined 
as $0.26\pm0.006$ mas after incorporating the
zero point offset (-0.05 mas) as suggested by \citet{2018ApJ...861..126R}.
The cluster distance corresponding to the average parallax is
obtained as $3.84\pm0.15$ kpc. This value is in good agreement with 
the distance estimated by \citet{2018A&A...618A..93C}.
 
We also used parallax of cluster stars along with the
proper motion data to find true cluster members.
A star is considered as probable member if it lies inside 
the circle of 0.7 mas $yr^{-1}$ in VPD and has
a parallax within $3\sigma$ from the mean cluster parallax. 
In this way we obtained a total 172 stars as probable cluster members for
King 13. We have matched our members with \citet{2018A&A...618A..93C} 
catalog as well. We have shown these matched stars with red open
circles in the observed colour magnitude and color-color diagrams. 
In our previous analysis  (\citet{2019MNRAS.490.1383R}), we
clearly explained this method in detail. The CMD of the probable cluster members having PM error $\le 1$ mas $yr^{-1}$ are
shown in the upper-middle panels of Fig \ref{pm_dist}. 
The main sequence of King 13 is clearly separated from the field stars.

\begin{figure}
\begin{center}
\hbox{
\includegraphics[width=6.5cm, height=6.5cm]{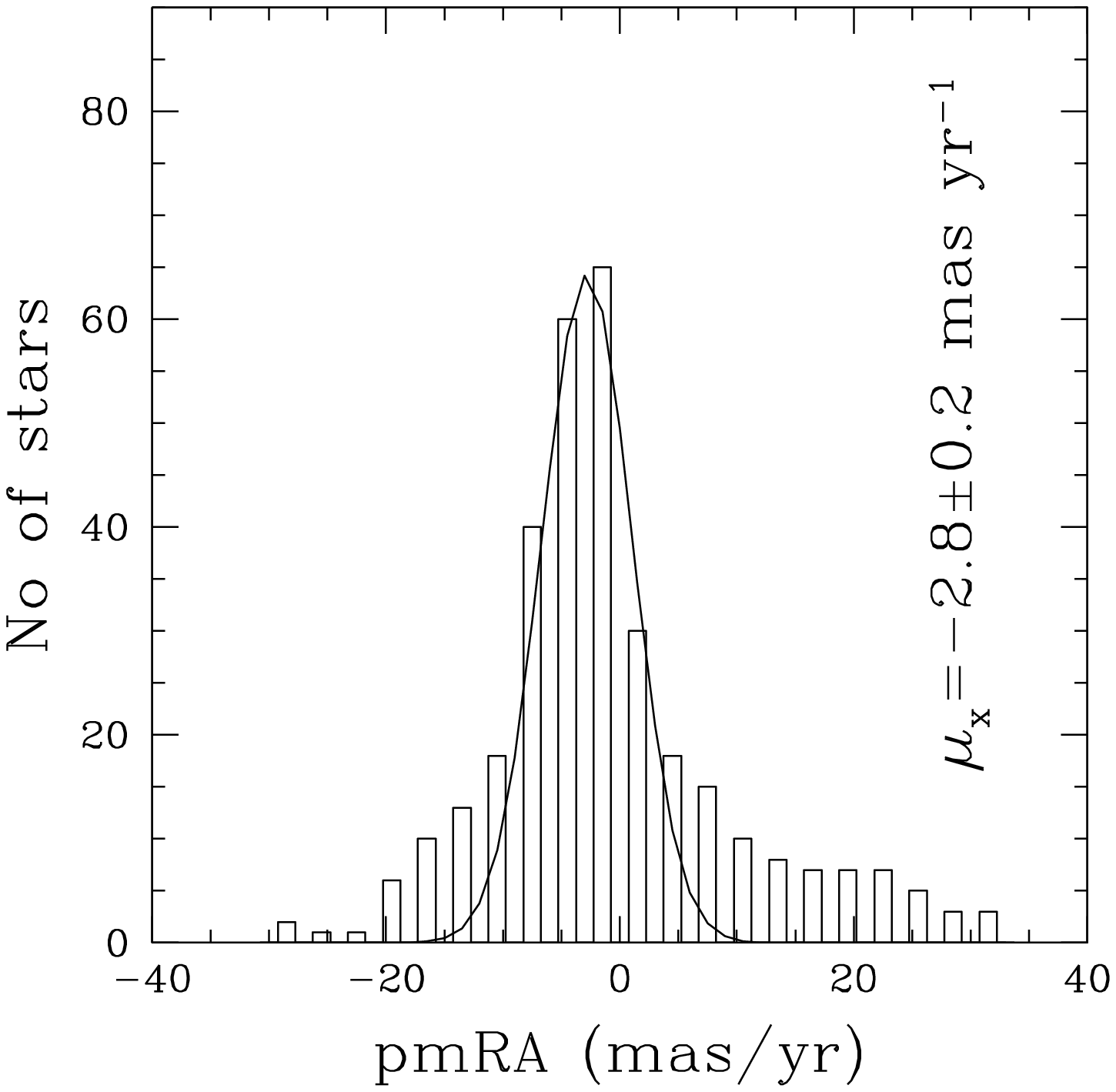}
\includegraphics[width=6.5cm, height=6.5cm]{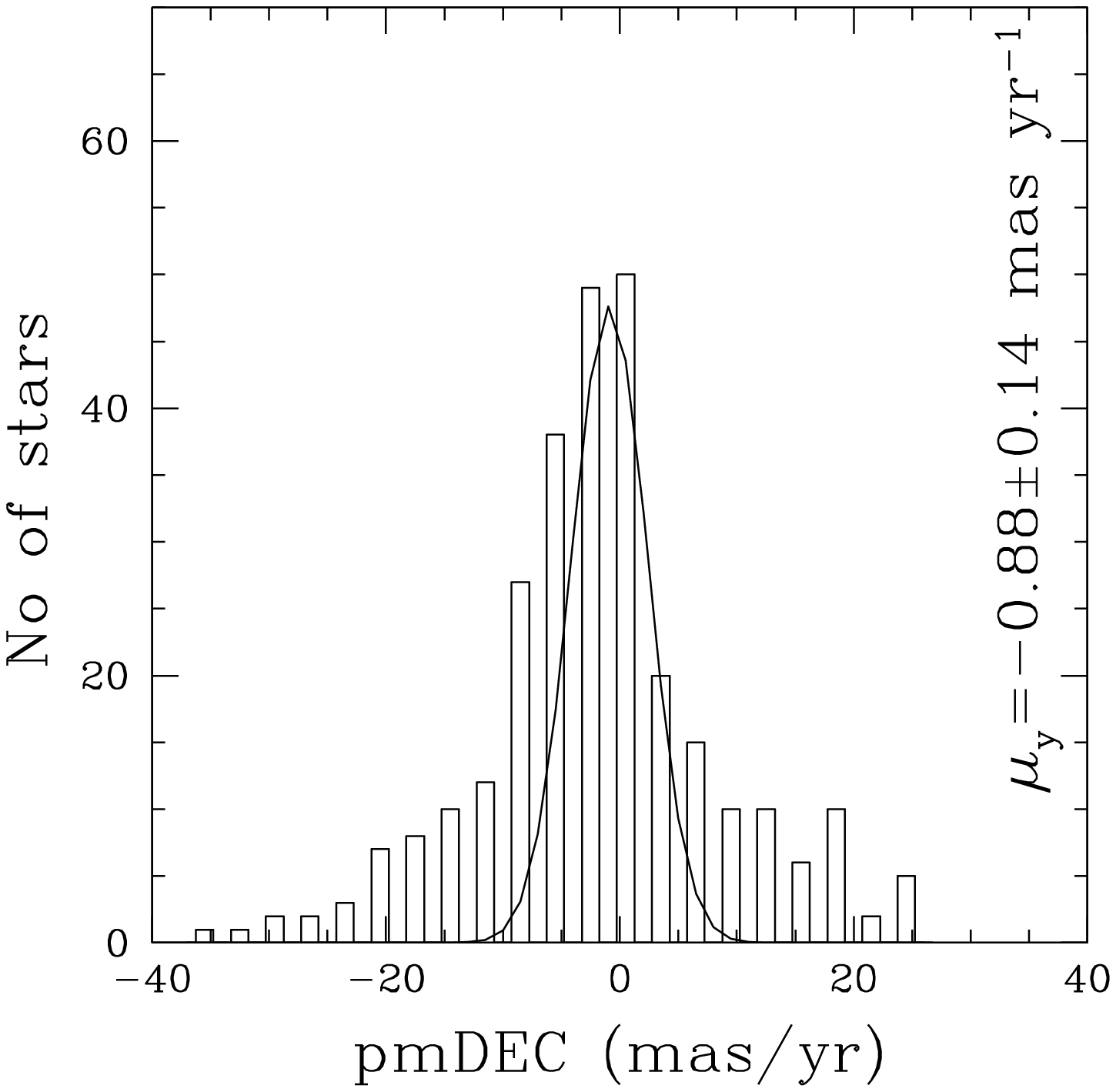}
}
\caption{The histograms for proper motion in right ascension (left) and declination (right).
The Gaussian function fit to the central bins provides the
mean values in RA and DEC.
panels.
 }
\label{pm}
\end{center}
\end{figure}

To calculate the mean proper motion of King 13, we constructed histograms of proper motions and fitted
their peaks with Gaussian functions ( see Fig \ref{pm}). We found $-2.8\pm0.2$ and $-0.9\pm0.1$ $mas$ $yr^{-1}$ as
mean proper motions in RA and DEC directions respectively. These proper motion values are very close to the values given
by \citet{2018A&A...618A..93C}. By visual inspection, we define a circle of 0.7 mas $yr^{-1}$ radius, around the cluster
center in the VPD which defines our member selection criteria. The chosen radius is a boundary between losing cluster members
with poor PMs and the inclusion of field stars. 

\section{Cluster structure and its fundamental parameters}
\label{ana}

\subsection{Cluster structure}

\begin{figure}
\begin{center}
\hbox{
\includegraphics[width=6.5cm, height=6.5cm]{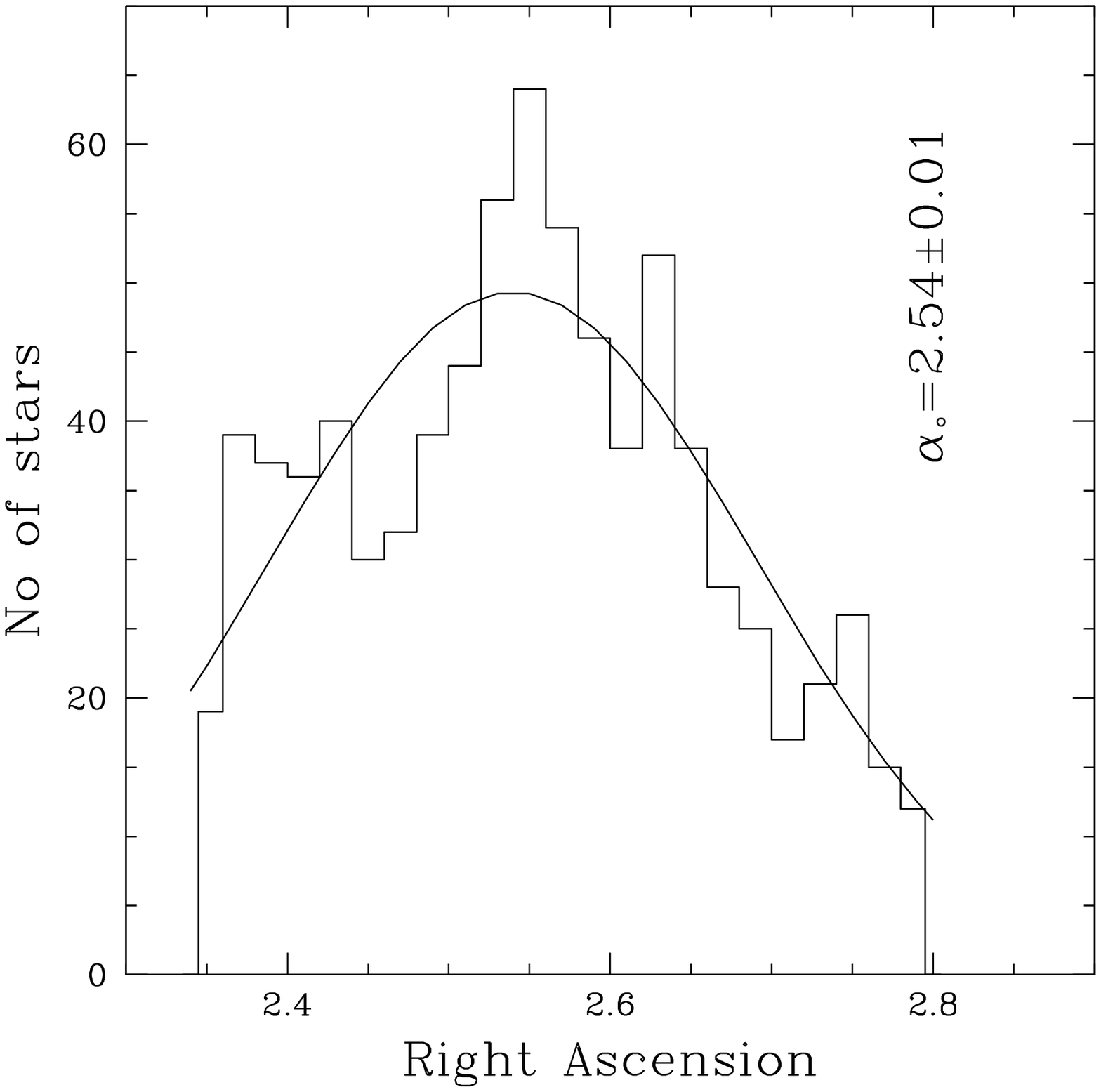}
\includegraphics[width=6.5cm, height=6.5cm]{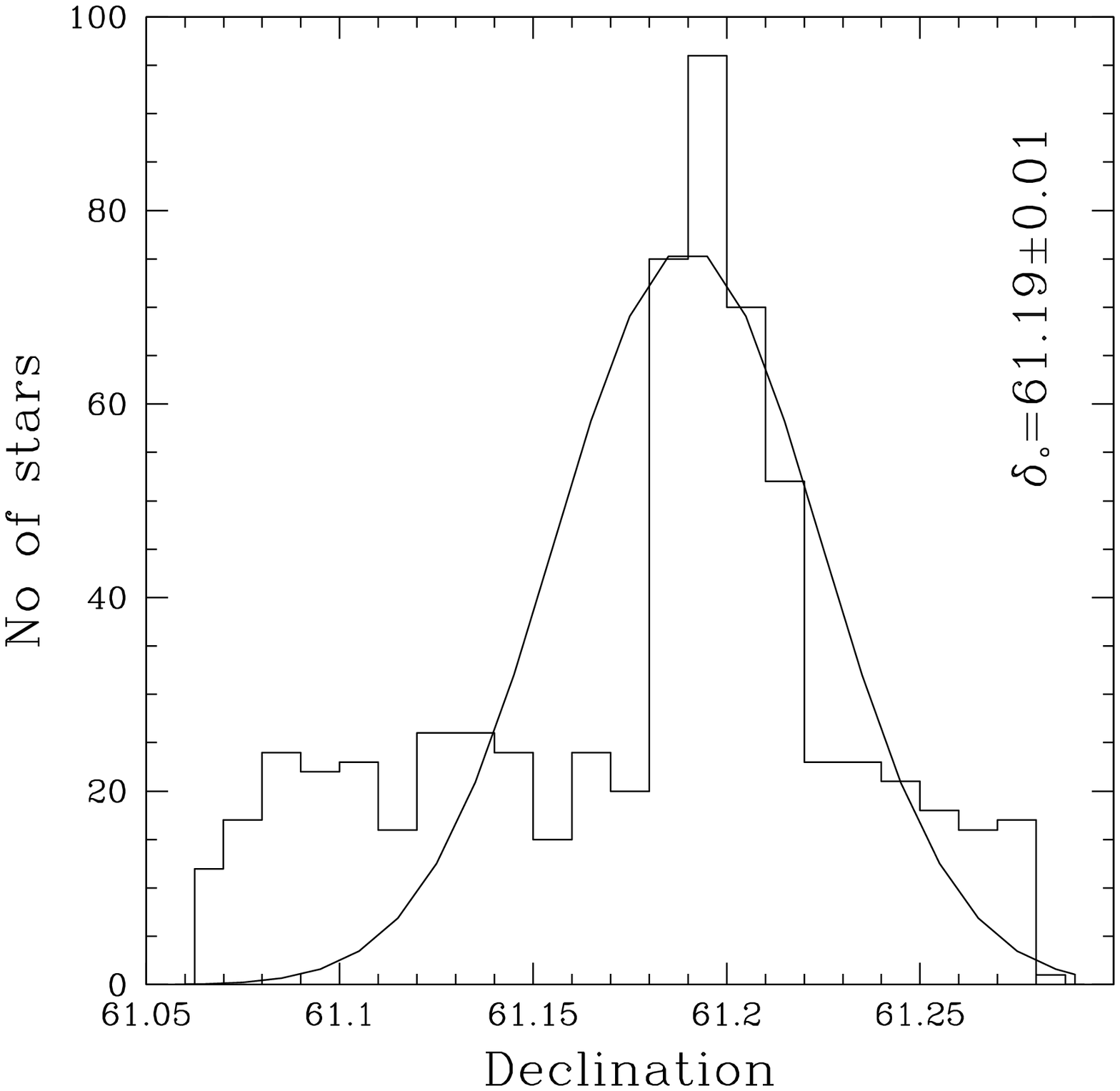}
}
\caption{The histograms for the estimation of center coordinates. The Gaussian function fit
to the central bins provides cluster center.}
\label{center}
\end{center}
\end{figure}

In order to investigate the cluster structure, the primary step is to find very precise center coordinates. For this, we
have plotted the histogram of star counts in Right Ascension (RA) and Declination (DEC) using Gaia database.
Our main aim is to estimate the maximum central density of the cluster. The cluster center is estimated by fitting Gaussian function
of star counts in RA and DEC as shown in Fig \ref{center}, finding coordinates as $\alpha=2.54\pm0.01$ deg
and $\delta=61.19\pm0.01$ deg. The cluster center in celestial coordinates is at $\alpha_{2000}$=00$^{h}$ 10$ ^{m}$
9.6$^{s}$, $\delta_{2000}$=61$^\circ$ 11$^\prime$ 24$^{\prime\prime}$. These coordinates are very close to the values given by
\citet{2010AstL...36...14G} and are also in good agreement with the center coordinates reported by \citet{2018A&A...618A..93C}.

\begin{figure}
\hspace{2cm}\includegraphics[width=10.5cm, height=10.5cm]{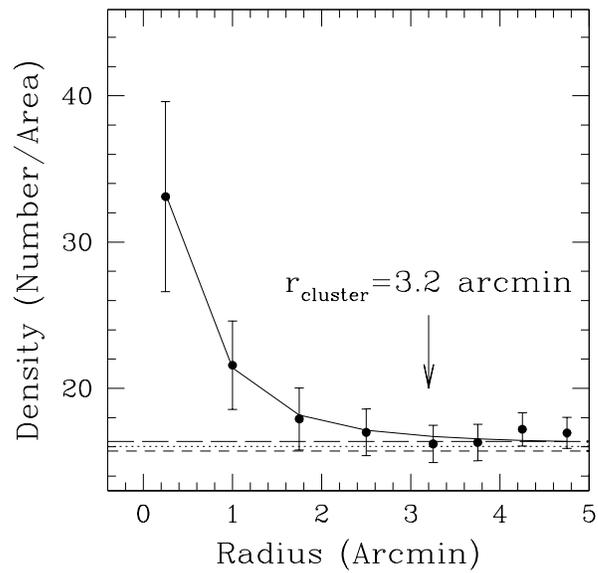}
\caption{Surface density distribution of stars in the field of the cluster King 13. Errors are
determined from sampling statistics(=$\frac{1}{\sqrt{N}}$ where $N$ is the number of stars used in
the density estimation at that point). The smooth line represent the fitted profile whereas dotted
line shows the background density level. Long and short dash lines represent the errors in background
density.}
\label{data}
\end{figure}

To estimate the cluster extent, we established the radial density profile (RDP) of King 13. To construct RDP the observed area 
of the cluster is divided into many concentric circles. The number density, $R_{i}$, in the i$^{th}$ zone is calculated by using
the formula $R_{i}$ = $\frac{N_{i}}{A_{i}}$, where $N_{i}$ is the number of stars and $A_{i}$ is the area of the i$^{th}$ zone.
RDP of the cluster King 13 is shown in Fig. \ref{data}. A smooth continuous line represents the fitted \citet{1962AJ.....67..471K}
profile, which can be expressed as:\\

~~~~~~~~~~~~~~~~~~~~~~~~~~$f(r) = f_{b}+\frac{f_{0}}{1+(r/r_{c})^2}$~~~~~~~~~~~~~~~~~~~(3)\\

where $f_0$ is the central density, $r_c$ is the core radius and $f_b$ is the background density. The background density level with
errors is shown with dotted lines in this figure. This RDP flattens at $r\sim$ 3.2 arcmin and begins to merge with the background
stellar density which can be seen in Fig. \ref{data}. Therefore, 
we consider 3.2 arcmin as the extent of this cluster. By fitting the King model to the cluster density profile, we estimated the
values of central density, background density and core radius as $36.3\pm4.2 stars/arcmin^{2}$, $16.04\pm0.3 stars/arcmin^{2}$ and
$0.6\pm0.1$ arcmin. These parameters are also listed in Table \ref{tab4}.

The density contrast parameter $\delta_{c} = 1 +\frac{f_{0}}{f_{b}}$ for the cluster is estimated as 2.8, which is smaller
than the values ($7\le \delta_{c}\le 23$) given for compact star clusters by \citet{2009MNRAS.397.1915B}. This shows that
King 13 is a sparse cluster.

The tidal radius generally depends on the effects of Galactic tidal fields and subsequent internal relaxation dynamical
evolution of clusters \citep{1988RMxAA..16...25A}. To evaluate the clusters tidal radius, we have adopted the following relation
as given by \citet{2001A&A...375..863J}\\

~~~~~~~~~~~~~~~~~~~~~~~~~~$R_{t} = 1.46\times (M_{c} )^{1/3}$~~~~~~~~~~~~~~~~~~~~(4)\\

where $R_{t}$ and $M_{c}$ are the tidal radius and the total mass (see Sec. \ref{mf}) of the cluster respectively. The estimated
value of the tidal radius is found as 8.5 pc.

\subsection{Interstellar extinction}

\begin{figure}
\begin{center}
\hbox{
\includegraphics[width=4.5cm, height=4.5cm]{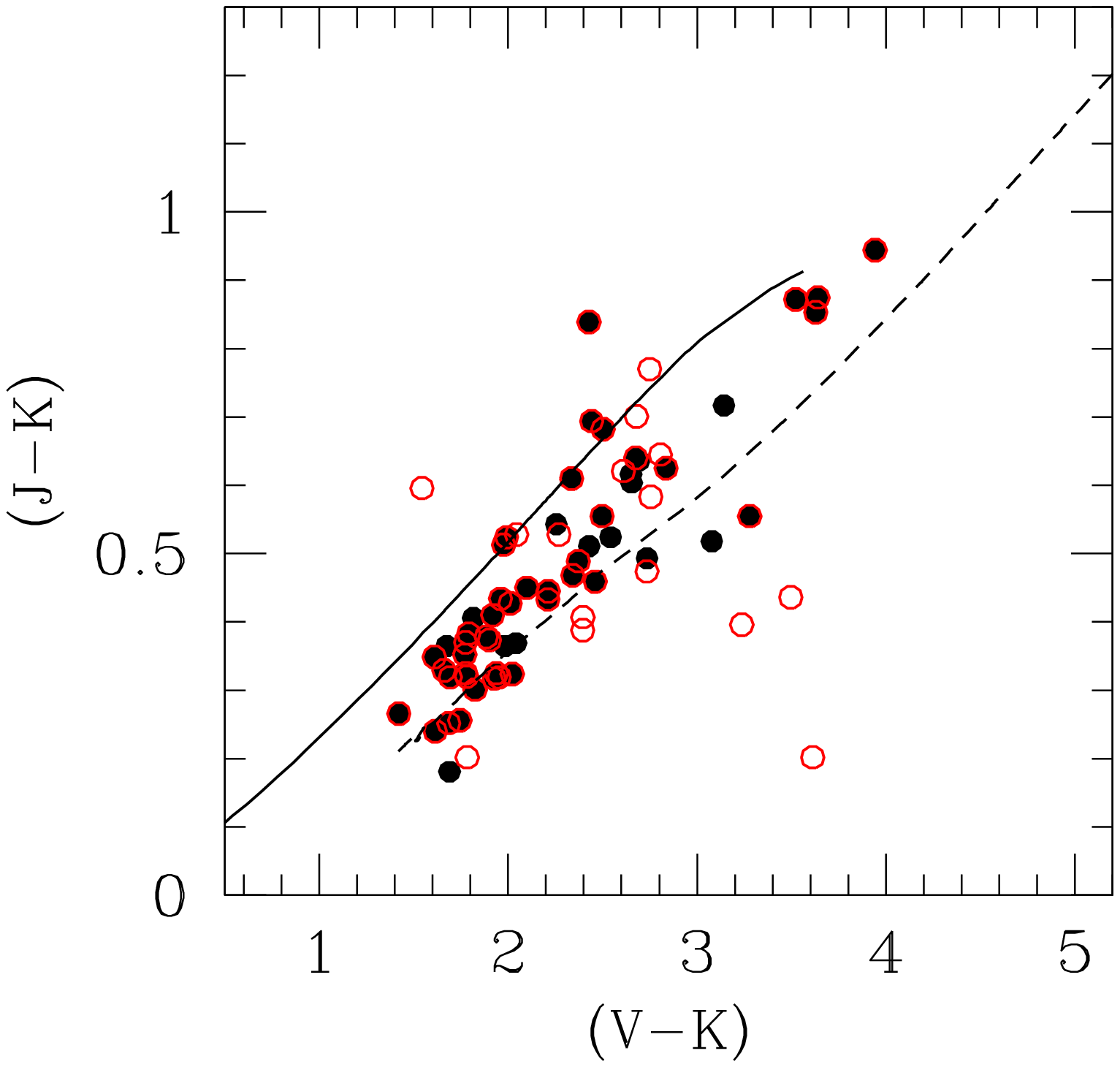}
\includegraphics[width=4.5cm, height=4.5cm]{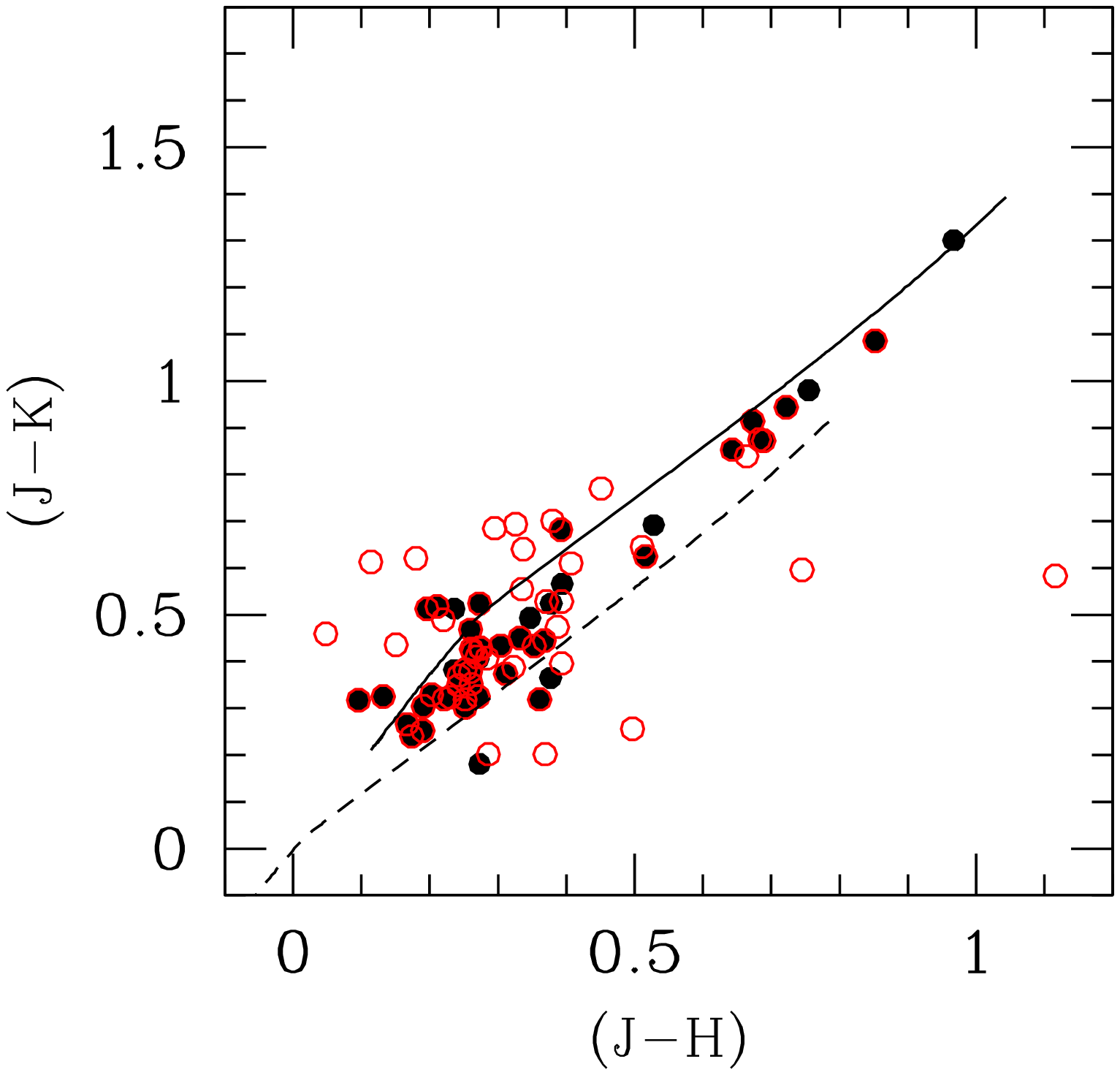}
\includegraphics[width=4.5cm, height=4.5cm]{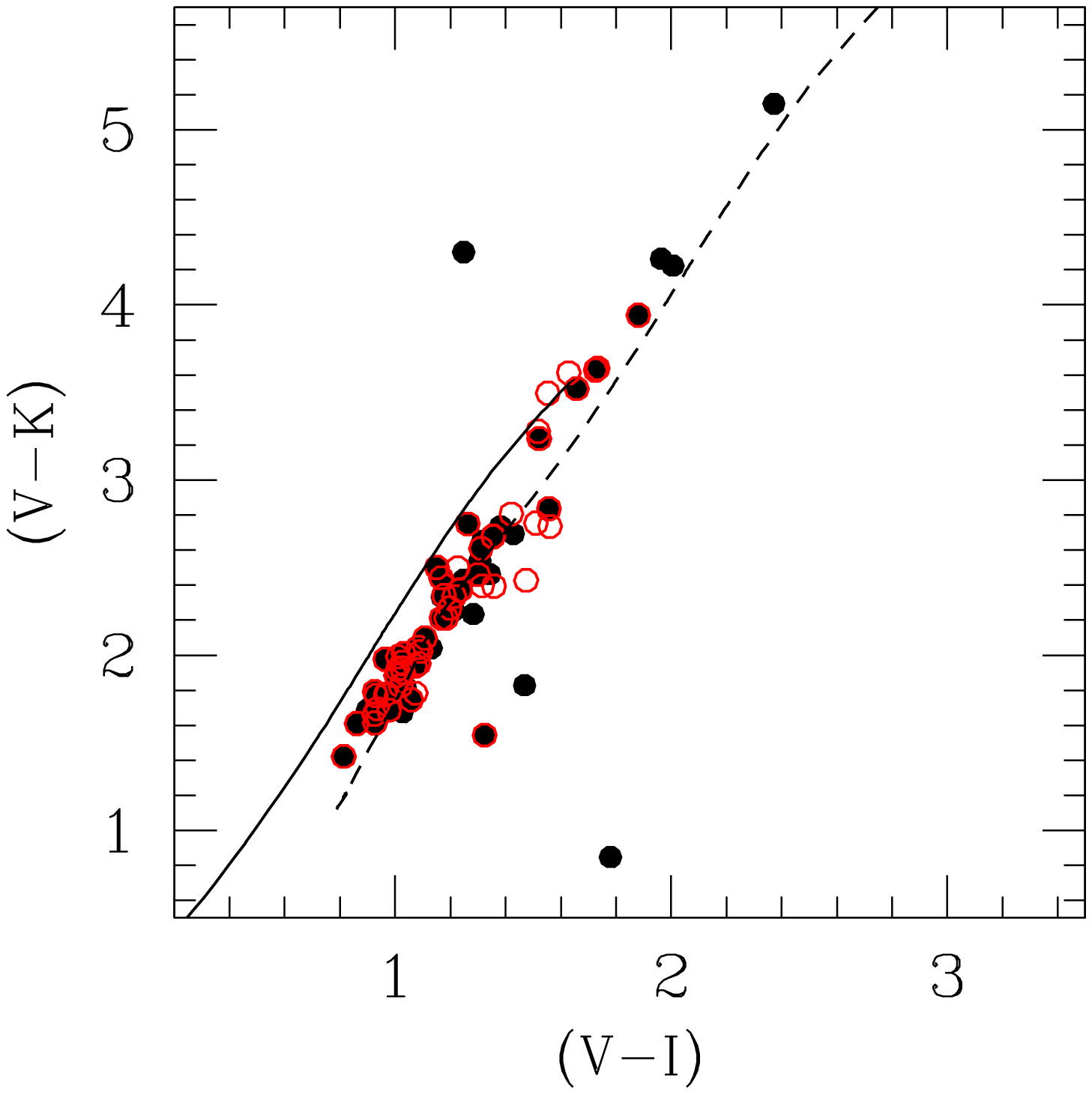}
}
\caption{The $(J-K), (V-K)$, $(J-K), (J-H)$ and $(V-I), (V-K)$ colour-colour diagrams for the cluster King 13.
The solid line is the ZAMS taken from \citet{2012MNRAS.427..127B}. The dotted line is the ZAMS shifted by the values given
in the text. Red open circles are matched stars with \citet{2018A&A...618A..93C} having membership probability higher than 0.5.}
\label{cc}
\end{center}
\end{figure}

The interstellar extinction and the ratio of total-to-selective extinction towards the cluster are very important for a proper
use of photometric data. The value of interstellar extinction $E(B-V)$ 
in the direction of King 13 has been estimated using near-IR $JHK$
data from 2MASS catalogue in combination with the optical data. We followed \citet{1998AJ....116.2475P} to convert $K_{s}$ magnitude
into $K$ magnitude. We plotted $(J-K)$ versus $(V-K)$, $(J-K)$ versus $(J-H)$ and  $(V-I)$ versus $(V-K)$ color-color diagrams which
are shown in Fig. \ref{cc}. The ZAMS by \citet{2012MNRAS.427..127B} is fitted over the color-color diagrams as shown by the solid black
line. We have also shown the shifted ZAMS also by dotted line in this figure. We have estimated $E(G_{BP}-G_{RP})=0.71\pm0.26$
and $A_{G}$=1.44 for King 13 using optical data from Gaia database in $G$, $G_{BP}$ and $G_{RP}$ filters. The error in the calculated value
of $E(G_{BP}-G_{RP})$ is the fitting error of ZAMS. With the help of this, errors in colour-excess are calculated in further transformations.
We used the absorption ratio in the optical and infrared wavelengths to visual absorption from \citet{1989ApJ...345..245C}. Using the 
transformation equations by \citet{2018JAsGe...7..180H}, we found the value of interstellar reddening $E(B-V)$ as $0.80\pm0.2$. The shifted
ZAMS provides $E(V-K) = 2.5$ mag, $E(J-K)=0.48\pm0.6$ mag and $E(J-H)=0.26\pm0.3$. We used the relations, $A_{k}=0.618\times 
E(J-K)$ \citep{1990ARA&A..28...37M}; $A_{k}=0.122\times A_{v}$ \citep{1989ApJ...345..245C} and $A_{v}=R\times E(B-V)$ to estimate the 
value of $A_{v}$ and $R$. Using the above relations we found the value of $A_{v}$ as 2.43 and $R$ as $\sim$3.04. Color excess 
ratios are calculated as $\frac{E(J-K)}{E(V-K)} \sim 0.19$ and $\frac{E(J-K)}{E(J-H)} \sim 1.84$, which are in good agreement with 
the normal interstellar extinction value as suggested by \citet{1989ApJ...345..245C}.

\subsection{Age and distance of the cluster}

The metallicity, age and distance of King 13 have been estimated by comparing the observed CMDs with theoretical stellar
evolutionary isochrones. For this purpose we adopted the isochrones of \citet{2012MNRAS.427..127B}. The main-sequence of the cluster
is well reproduced by isochrones with a nearly solar metallicity, Z=0.012, which have been adopted for the following analysis.

\begin{figure*}
\begin{center}
\hbox{
\includegraphics[width=8.5cm,height=8.5cm]{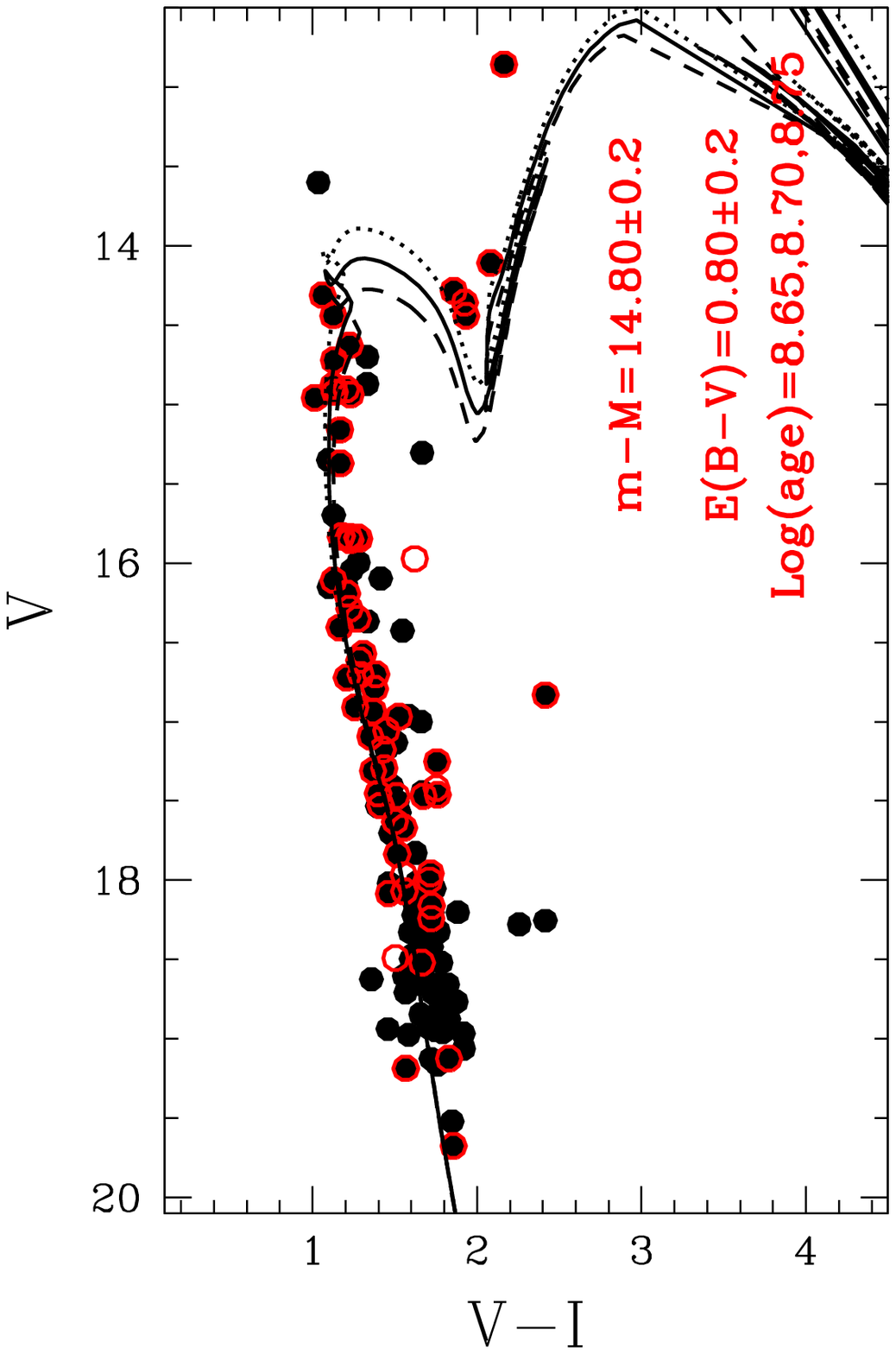}
\includegraphics[width=8.5cm,height=8.5cm]{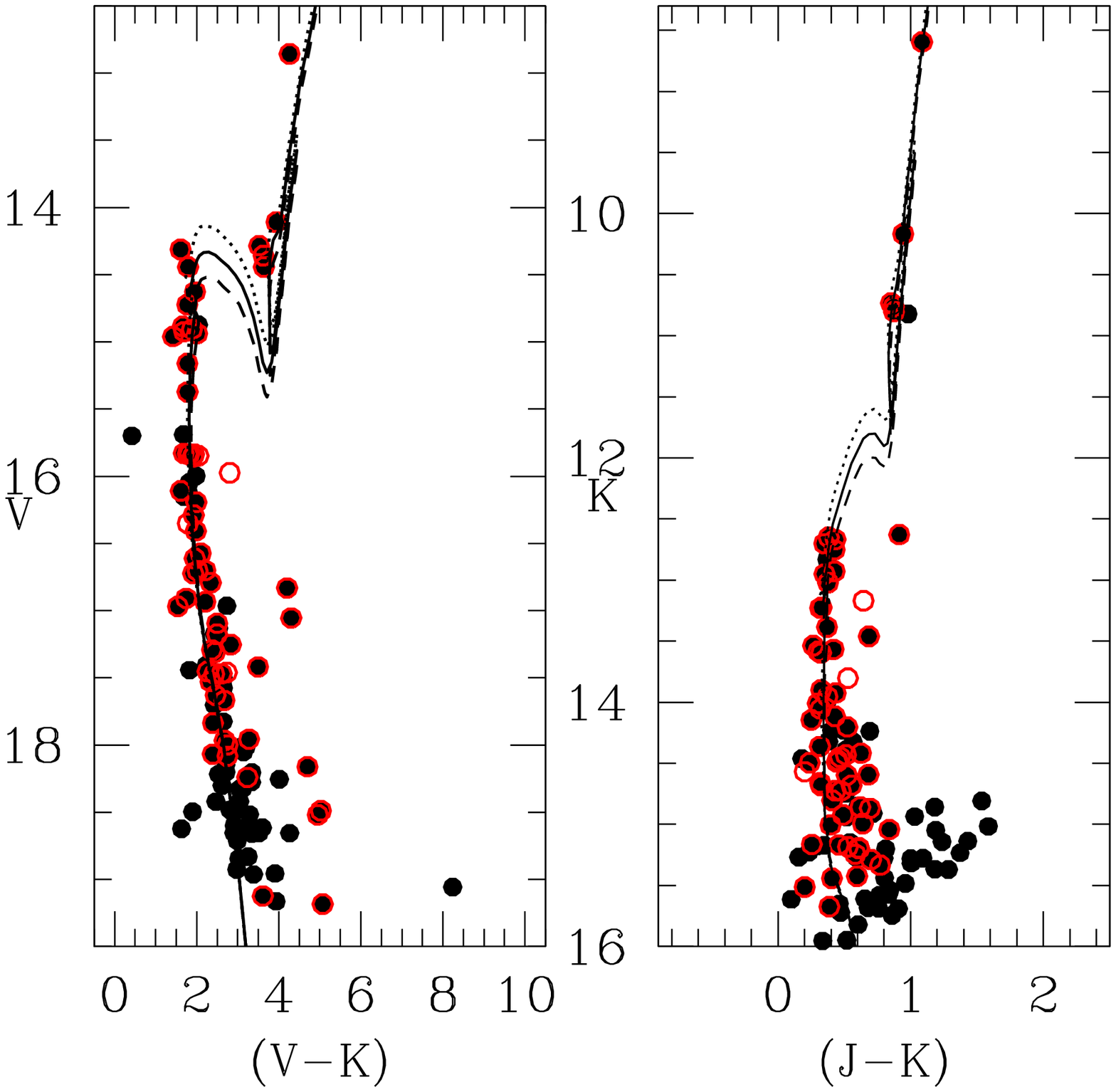}
}
\caption{ The $V$, $(B-V)$, $V$, $(V-K)$ and $K$, $(J-K)$ colour-magnitude diagram of the cluster 
        King 13. The curves are the isochrones of (log(age) $=$  8.65 ,8.70 and 8.75). These ishochrones
        are taken from  \citet{2012MNRAS.427..127B}. Red open circles are matched stars with \citet{2018A&A...618A..93C}
        having membership probability higher than 0.5.}
\label{cmd_vi}
\end{center}
\end{figure*}

We surveyed different age isochrones to get the best fit with morphological features in $V, ~(V-I)$; $V, ~(V-K)$
and $K, ~(J-K)$ CMDs. To get a clear sequence in the CMDs, we consider only probable cluster members based on
the cluster's VPD. In Fig. \ref{cmd_vi}, we superimpose isochrones of different age (log(age)=8.65, 8.70 and 8.75) having
$Z=0.012$ in $V, ~(V-I)$; $V, ~(V-K)$ and $K, ~(J-K)$ CMDs. Assuming the brightest star as an evolved star, we found an age
of $510\pm60$ Myr. The distance modulus for the cluster is found to be $(m-M)=14.8\pm0.2$ mag which corresponds to a 
heliocentric distance of $3.75\pm0.4$ kpc. Present estimated distance of King 13 is 2.1 kpc higher than 
the previously derived value by \citet{2014Ap&SS.352..665H} using 2MASS $JHKs$ data. The Galactocentric coordinates for the cluster are
determined as $X=3.68$ kpc, $Y=9.57$ kpc and $Z=-0.08$ kpc. The calculated $Z$ value indicates that the cluster King 13
is in the thin Galactic disk. The Galactocentric distance of the cluster is determined as $11.23$ kpc.

\section{Mass function and dynamical state of King 13}
\label{mf}

\subsection{Luminosity function}

The distribution of the stars of an OC based on their brightness is termed as 
luminosity function. To get the luminosity function, it is very important to know
the completeness factor (CF) of CCD data. We implemented the artificial star test 
for the estimation of CF. To perform this test we have used ADDSTAR routine in DAOPHOT II. We randomly added many stars in different
magnitude bins to the original $V$ and $I$ images having the same geometrical positions. We added $\sim$ 10 $\%$ of the number of stars actually
detected and inserted more stars into fainter magnitude bins. Then we again performed the same procedure of photometry for new images as well.
The CF is calculated as the ratio between the number of artificial stars recovered in $V$ and $I$ passbands and the number of added stars
per magnitude bin. The values of CF are listed in Table \ref{comp} corresponding to each $V$ mag bin. From this table, we can conclude that
almost every star has been recovered in the brighter end and as we go towards the fainter end the completeness of the data decreases.

\begin{figure}
\centering
\includegraphics[width=6.5cm, height=6.5cm]{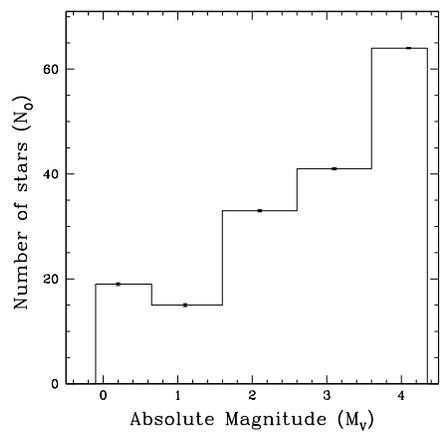}
\vspace{-0.5cm}
\caption{The luminosity functions of the cluster under consideration.}
\label{lf}
\end{figure}

To establish the luminosity function for King 13, we have used $V$ versus $(V-I)$ CMD. Firstly, we converted the apparent $V$ magnitudes
of probable member stars into the absolute magnitudes using the value of distance modulus. To remove field star contamination completely
from the main sequence of King 13, we used probable cluster 
members selected by using vector point diagram and parallax. 
Then we constructed the true histogram of
LF as shown in Fig. \ref{lf}. The histogram shows that the luminosity function for the cluster King 13 rises steadily up to 4.2 mag.


\subsection{Mass function}

\begin{figure}
\centering
\includegraphics[width=6.5cm, height=6.5cm]{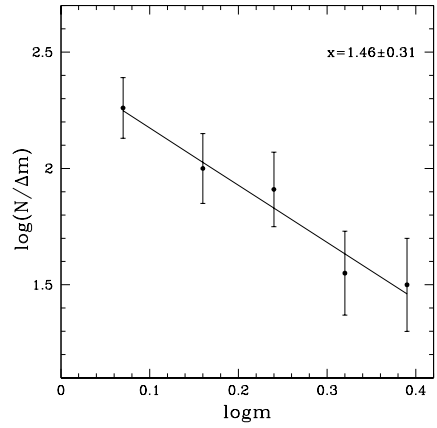}
\vspace{-0.5cm}
\caption{Mass function for King 13 derived using \citet{2012MNRAS.427..127B} isochrones. Standard deviations from
the central values are represented by the error bars.}
\label{mass}
\end{figure}

For the main sequence stars, the LF is transformed into the mass function (MF) using the theoretical 
model given by \citet{2012MNRAS.427..127B}. The resulting mass function is plotted in Fig \ref{mass}.

The mass function slope has been derived by using the relation $\log\frac{dN}{dM}=-(1+x)\log(M)$+constant,
where $dN$ is the number of stars in a mass bin $dM$ with central mass $M$ and $x$ is the slope of MF. The derived
MF slope ($x= 1.46\pm0.31$) is more precise than \citet{2014Ap&SS.352..665H} because the current analysis is based on
deep CCD $VI$ data along with GAIA~DR2 astrometry. Our derived value of MF slope for this cluster in the mass range
1.1-2.6$M_{\odot}$ is in good agreement with the value 1.35 given by \citet{1955ApJ...121..161S} for field stars in the Solar
neighbourhood.

The Total mass of King 13 was estimated as $\sim$270 $M_{\odot}$, considering the derived mass function slope within the mass range
1.1$M_{\odot}$~-~2.6$M_{\odot}$ and using the following relationship as
used by \citet{2008MNRAS.390..985Y}  \\

~~~~~~~~~~~~~~~~~~~~~~~~~~$M = C \int_{M_{L}}^{M_{U}} M^{\Gamma}M dM$~~~~~~~~~~~~~~~~~~~~~(5)     \\

where C is the constant, $M_{L}$ and $M_{U}$ are the upper and lower
mass limits of the cluster stars, $\Gamma$ is the slope of the mass
function.

\subsection{Dynamical state of the cluster}

\begin{figure}
\centering
\includegraphics[width=6.5cm, height=6.5cm]{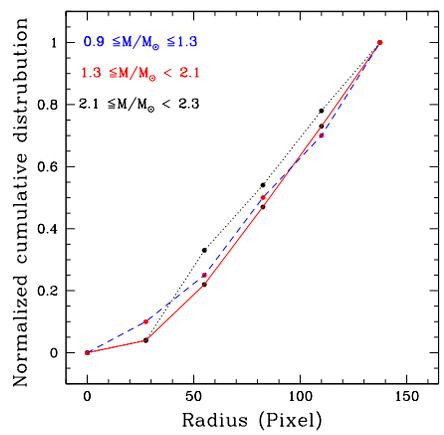}
\vspace{-0.5cm}
\caption{The cumulative radial distribution of stars in various mass range.}
\label{mass_seg}
\end{figure}

To characterize the degree of mass-segregation effect in King 13,  we plotted the cumulative radial distribution (CRD) of
stars for various mass ranges in Fig. \ref{mass_seg}. The main sequence stars are subdivided into three mass ranges  
2.1$\le\frac{M}{M_{\odot}}<$~2.3, 1.3$\le\frac{M}{M_{\odot}}<$~2.1 and 0.9$\le\frac{M}{M_{\odot}}\le$~1.3.
Fig. \ref{mass_seg} indicates that more massive stars tend to lie toward the cluster center. The Kolmogorov-Smirnov
test $(K-S)$ test shows the evidence for statistical significance of this effect with a confidence level of 80 \%. CRD of
the stars having a mass range  $2.1<\frac{M}{M_{\odot}} \le 2.3$ can be discriminated from the CRD of fainter stars having mass range
$0.9 \le \frac{M}{M_{\odot}} \le 1.3$ at a confidence level of 80 \%, which clearly shows the evidence of
mass-segregation.

Mass segregation effect in clusters may be attributed to dynamical evolution or star formation or both. During the lifetime of a
cluster, encounters between its members moderately lead to an increased degree of energy equipartition. At the same time, bright
stars gradually sink towards the cluster center and deliver their kinetic energy to the low mass stars, thus leading to this effect.
The relaxation time $T_{E}$, is the time scale in which the cluster will lose the memory of dynamical initial conditions.
Mathematically, $T_{E}$ is denoted by the following formula \citep{1971ApJ...164..399S}:\\

~~~~~~~~~~~~~~~~~~~~~~~~~~~~~~$T_{E}=\frac{8.9\times10^5\sqrt{N}\times{R_{h}}^{3/2}}{\sqrt{m}\times log(0.4N)}$~~~~~~~~~~~~~~~~~~~~~~~~~(6) \\

where $N$ is the number of cluster members, $R_{h}$ is the half mass radius of the cluster and $<$m$>$ is the mean mass
of the cluster stars \citep{1971ApJ...164..399S}. The value of $R_{h}$ has been assumed as half of the cluster extent
value derived by us. Using the above formula, we have estimated the dynamical relaxation time of King 13 as 7.5 Myr.

The estimated values of relaxation time for this cluster is less than the cluster age. Therefore we conclude that
King 13 is a dynamically relaxed cluster.

\section{Conclusion}
\label{con}

We present a $VI$ $CCD$ photometric, 2MASS near-IR and Gaia DR2 astrometric study of the open cluster King 13, which
is not well studied in the literature. The estimated fundamental parameters are listed in Table \ref{tab5}. The main results
of the present analysis are summarized as follows:

\begin{enumerate}

\item  To separate cluster members from the field stars, we used 
Gaia~DR2 proper motion and parallax data and obtained a clear
main sequence for the cluster. The mean proper motion of the cluster is determined as $-2.8 \pm 0.2$ and $-0.9 \pm 0.1$ mas
$yr^{-1}$ in RA and DEC directions respectively.

\item  We calculated the structural properties of the cluster. The Center of the cluster is determined as
$\alpha = 00^{h} 10^{m} 9.6^{s}$, $\delta_{2000} = 61^{\circ} 11^{\prime} 24^{\prime \prime}$. The Radius
and tidal radius of the cluster are determined as $3.2^{\prime}$ and 8.5 pc respectively. The value of density
contrast parameter $\delta_{c}$ shows that King 13 is a sparse cluster.

\item  The values of color-excess ratios $\frac{E(J-K)}{E(V-K)}$ and $\frac{E(J-K)}{E(J-H)}$ are found to
be $\sim 0.19$ and  1.84, which show a normal interstellar extinction in the direction of King 13.

\item  From the comparison of the cluster CMDs with theoretical stellar evolutionary isochrones with a metallicity Z=0.012
(\citet{2012MNRAS.427..127B}), we determined age, reddening and distance of the cluster as $510 \pm 60$ Myr, $0.80 \pm 0.2$ mag and
$3.84 \pm 0.15$ kpc respectively.

\item  To study the dynamical properties of cluster, we constructed luminosity and mass function for the cluster.
The luminosity function increases towards the fainter end. This indicates that fainter stars are still bound to
the cluster. Mass function slope for the cluster is determined as $1.46 \pm 0.31$ which is in agreement with
\citet{1955ApJ...121..161S} value. The cumulative radial stellar distribution plot shows presence of mass segregation in the
cluster. Dynamical relaxation time of the cluster is found to be 7.5 Myr, which shows that King 13 is a dynamically
relaxed cluster.

\end{enumerate}

\section{ACKNOWLEDGEMENTS}
\label{ACK}

We thank the staff of ARIES for assistance during observations and data reduction. This work has been partially
supported by the Natural Science Foundation of China (NSFC-11590782, NSFC-11421303). This work has made use of data
from the European Space Agency (ESA) mission GAIA processed by 
Gaia Data processing and Analysis Consortium (DPAC,
$https://www.cosmos.esa.int/web/gaia/dpac/consortium$). 
This publication has made use of data from the 2MASS, which
is a joint project of the University of Massachusetts and the Infrared Processing and Analysis Center/California
Institute of Technology, funded by the National Science Foundation.


  \bibliographystyle{elsarticle-harv}
  \bibliography{ms_revised}

\begin{thebibliography}{34}
\expandafter\ifx\csname natexlab\endcsname\relax\def\natexlab#1{#1}\fi
\providecommand{\url}[1]{\texttt{#1}}
\providecommand{\href}[2]{#2}
\providecommand{\path}[1]{#1}
\providecommand{\DOIprefix}{doi:}
\providecommand{\ArXivprefix}{arXiv:}
\providecommand{\URLprefix}{URL: }
\providecommand{\Pubmedprefix}{pmid:}
\providecommand{\doi}[1]{\href{http://dx.doi.org/#1}{\path{#1}}}
\providecommand{\Pubmed}[1]{\href{pmid:#1}{\path{#1}}}
\providecommand{\bibinfo}[2]{#2}
\ifx\xfnm\relax \def\xfnm[#1]{\unskip,\space#1}\fi
\bibitem[{{Allen} and {Martos}(1988)}]{1988RMxAA..16...25A}
\bibinfo{author}{{Allen}, C.}, \bibinfo{author}{{Martos}, M.A.},
  \bibinfo{year}{1988}.
\newblock \bibinfo{title}{{The galactic orbits and tidal radii of selected star
  clusters.}}
\newblock \bibinfo{journal}{\rmxaa} \bibinfo{volume}{16},
  \bibinfo{pages}{25--36}.
\bibitem[{{Bisht} et~al.(2016){Bisht}, {Yadav} and
  {Durgapal}}]{2016NewA...42...66B}
\bibinfo{author}{{Bisht}, D.}, \bibinfo{author}{{Yadav}, R.K.S.},
  \bibinfo{author}{{Durgapal}, A.K.}, \bibinfo{year}{2016}.
\newblock \bibinfo{title}{{Photometric study of open star clusters in II
  quadrant: Teutsch 1 and Riddle 4}}.
\newblock \bibinfo{journal}{\na} \bibinfo{volume}{42}, \bibinfo{pages}{66--77}.
\newblock \DOIprefix\doi{10.1016/j.newast.2015.06.005}.
\bibitem[{{Bisht} et~al.(2019){Bisht}, {Yadav}, {Ganesh}, {Durgapal}, {Rangwal}
  and {Fynbo}}]{2019MNRAS.482.1471B}
\bibinfo{author}{{Bisht}, D.}, \bibinfo{author}{{Yadav}, R.K.S.},
  \bibinfo{author}{{Ganesh}, S.}, \bibinfo{author}{{Durgapal}, A.K.},
  \bibinfo{author}{{Rangwal}, G.}, \bibinfo{author}{{Fynbo}, J.P.U.},
  \bibinfo{year}{2019}.
\newblock \bibinfo{title}{{Mass function and dynamical study of the open
  clusters Berkeley 24 and Czernik 27 using ground based imaging and Gaia
  astrometry}}.
\newblock \bibinfo{journal}{\mnras} \bibinfo{volume}{482},
  \bibinfo{pages}{1471--1484}.
\newblock \DOIprefix\doi{10.1093/mnras/sty2781},
  \href{http://arxiv.org/abs/1810.05380}{{\tt arXiv:1810.05380}}.
\bibitem[{{Bonatto} and {Bica}(2009)}]{2009MNRAS.397.1915B}
\bibinfo{author}{{Bonatto}, C.}, \bibinfo{author}{{Bica}, E.},
  \bibinfo{year}{2009}.
\newblock \bibinfo{title}{{The nature of the young and low-mass open clusters
  Pismis5, vdB80, NGC1931 and BDSB96}}.
\newblock \bibinfo{journal}{\mnras} \bibinfo{volume}{397},
  \bibinfo{pages}{1915--1925}.
\newblock \DOIprefix\doi{10.1111/j.1365-2966.2009.14877.x},
  \href{http://arxiv.org/abs/0904.1321}{{\tt arXiv:0904.1321}}.
\bibitem[{{Bressan} et~al.(2012){Bressan}, {Marigo}, {Girardi}, {Salasnich},
  {Dal Cero}, {Rubele} and {Nanni}}]{2012MNRAS.427..127B}
\bibinfo{author}{{Bressan}, A.}, \bibinfo{author}{{Marigo}, P.},
  \bibinfo{author}{{Girardi}, L.}, \bibinfo{author}{{Salasnich}, B.},
  \bibinfo{author}{{Dal Cero}, C.}, \bibinfo{author}{{Rubele}, S.},
  \bibinfo{author}{{Nanni}, A.}, \bibinfo{year}{2012}.
\newblock \bibinfo{title}{{PARSEC: stellar tracks and isochrones with the
  PAdova and TRieste Stellar Evolution Code.}}
\newblock \bibinfo{journal}{\mnras} \bibinfo{volume}{427},
  \bibinfo{pages}{127--145}.
\bibitem[{{Bukowiecki} et~al.(2011){Bukowiecki}, {Maciejewski}, {Konorski} and
  {Strobel}}]{2011AcA....61..231B}
\bibinfo{author}{{Bukowiecki}, {\L}.}, \bibinfo{author}{{Maciejewski}, G.},
  \bibinfo{author}{{Konorski}, P.}, \bibinfo{author}{{Strobel}, A.},
  \bibinfo{year}{2011}.
\newblock \bibinfo{title}{{Open Clusters in 2MASS Photometry. I. Structural and
  Basic Astrophysical Parameters}}.
\newblock \bibinfo{journal}{\actaa} \bibinfo{volume}{61},
  \bibinfo{pages}{231--246}.
\newblock \href{http://arxiv.org/abs/1107.5119}{{\tt arXiv:1107.5119}}.
\bibitem[{{Cantat-Gaudin} et~al.(2018){Cantat-Gaudin}, {Jordi}, {Vallenari},
  {Bragaglia}, {Balaguer-N{\'u}{\~n}ez}, {Soubiran}, {Bossini}, {Moitinho},
  {Castro-Ginard}, {Krone-Martins}, {Casamiquela}, {Sordo} and
  {Carrera}}]{2018A&A...618A..93C}
\bibinfo{author}{{Cantat-Gaudin}, T.}, \bibinfo{author}{{Jordi}, C.},
  \bibinfo{author}{{Vallenari}, A.}, \bibinfo{author}{{Bragaglia}, A.},
  \bibinfo{author}{{Balaguer-N{\'u}{\~n}ez}, L.}, \bibinfo{author}{{Soubiran},
  C.}, \bibinfo{author}{{Bossini}, D.}, \bibinfo{author}{{Moitinho}, A.},
  \bibinfo{author}{{Castro-Ginard}, A.}, \bibinfo{author}{{Krone-Martins}, A.},
  \bibinfo{author}{{Casamiquela}, L.}, \bibinfo{author}{{Sordo}, R.},
  \bibinfo{author}{{Carrera}, R.}, \bibinfo{year}{2018}.
\newblock \bibinfo{title}{{A Gaia DR2 view of the open cluster population in
  the Milky Way}}.
\newblock \bibinfo{journal}{\aap} \bibinfo{volume}{618}.
\newblock \DOIprefix\doi{10.1051/0004-6361/201833476}.
\bibitem[{{Cardelli} et~al.(1989){Cardelli}, {Clayton} and
  {Mathis}}]{1989ApJ...345..245C}
\bibinfo{author}{{Cardelli}, J.A.}, \bibinfo{author}{{Clayton}, G.C.},
  \bibinfo{author}{{Mathis}, J.S.}, \bibinfo{year}{1989}.
\newblock \bibinfo{title}{{The relationship between infrared, optical, and
  ultraviolet extinction}}.
\newblock \bibinfo{journal}{\apj} \bibinfo{volume}{345},
  \bibinfo{pages}{245--256}.
\newblock \DOIprefix\doi{10.1086/167900}.
\bibitem[{{Carraro} et~al.(2010){Carraro}, {Costa} and
  {Ahumada}}]{2010AJ....140..954C}
\bibinfo{author}{{Carraro}, G.}, \bibinfo{author}{{Costa}, E.},
  \bibinfo{author}{{Ahumada}, J.A.}, \bibinfo{year}{2010}.
\newblock \bibinfo{title}{{Photometric Characterization of the Galactic Star
  Cluster Trumpler 20}}.
\newblock \bibinfo{journal}{\aj} \bibinfo{volume}{140},
  \bibinfo{pages}{954--961}.
\newblock \DOIprefix\doi{10.1088/0004-6256/140/4/954},
  \href{http://arxiv.org/abs/1007.4782}{{\tt arXiv:1007.4782}}.
\bibitem[{{Durgapal} and {Pandey}(2001)}]{2001A&A...375..840D}
\bibinfo{author}{{Durgapal}, A.K.}, \bibinfo{author}{{Pandey}, A.K.},
  \bibinfo{year}{2001}.
\newblock \bibinfo{title}{{Structure and mass function of five intermediate/old
  open clusters}}.
\newblock \bibinfo{journal}{\aap} \bibinfo{volume}{375},
  \bibinfo{pages}{840--850}.
\newblock \DOIprefix\doi{10.1051/0004-6361:20010892}.
\bibitem[{{Durgapal} et~al.(1997){Durgapal}, {Pandey} and
  {Mohan}}]{1997BASI...25..489D}
\bibinfo{author}{{Durgapal}, A.K.}, \bibinfo{author}{{Pandey}, A.K.},
  \bibinfo{author}{{Mohan}, V.}, \bibinfo{year}{1997}.
\newblock \bibinfo{title}{{CCD photometry of galactic open star clusters - V.
  King 7}}.
\newblock \bibinfo{journal}{Bulletin of the Astronomical Society of India}
  \bibinfo{volume}{25}, \bibinfo{pages}{489}.
\bibitem[{{Friel}(1995)}]{1995ARA&A..33..381F}
\bibinfo{author}{{Friel}, E.D.}, \bibinfo{year}{1995}.
\newblock \bibinfo{title}{{The Old Open Clusters Of The Milky Way}}.
\newblock \bibinfo{journal}{\araa} \bibinfo{volume}{33},
  \bibinfo{pages}{381--414}.
\newblock \DOIprefix\doi{10.1146/annurev.aa.33.090195.002121}.
\bibitem[{{Gaia Collaboration} et~al.(2016){Gaia Collaboration}, {Prusti}, {de
  Bruijne}, {Brown}, {Vallenari}, {Babusiaux}, {Bailer-Jones}, {Bastian},
  {Biermann}, {Evans} and et~al.}]{2016A&A...595A...1G}
\bibinfo{author}{{Gaia Collaboration}}, \bibinfo{author}{{Prusti}, T.},
  \bibinfo{author}{{de Bruijne}, J.H.J.}, \bibinfo{author}{{Brown}, A.G.A.},
  \bibinfo{author}{{Vallenari}, A.}, \bibinfo{author}{{Babusiaux}, C.},
  \bibinfo{author}{{Bailer-Jones}, C.A.L.}, \bibinfo{author}{{Bastian}, U.},
  \bibinfo{author}{{Biermann}, M.}, \bibinfo{author}{{Evans}, D.W.},
  \bibinfo{author}{et~al.}, \bibinfo{year}{2016}.
\newblock \bibinfo{title}{{The Gaia mission}}.
\newblock \bibinfo{journal}{\aap} \bibinfo{volume}{595}, \bibinfo{pages}{A1}.
\newblock \DOIprefix\doi{10.1051/0004-6361/201629272},
  \href{http://arxiv.org/abs/1609.04153}{{\tt arXiv:1609.04153}}.
\bibitem[{{Glushkova} et~al.(2010){Glushkova}, {Zabolotskikh}, {Koposov},
  {Spiridonova}, {Vlasyuk} and {Rastorguev}}]{2010AstL...36...14G}
\bibinfo{author}{{Glushkova}, E.V.}, \bibinfo{author}{{Zabolotskikh}, M.V.},
  \bibinfo{author}{{Koposov}, S.E.}, \bibinfo{author}{{Spiridonova}, O.I.},
  \bibinfo{author}{{Vlasyuk}, V.V.}, \bibinfo{author}{{Rastorguev}, A.S.},
  \bibinfo{year}{2010}.
\newblock \bibinfo{title}{{Photometry of the poorly studied galactic open star
  clusters King 13, King 18, King 19, King 20, NGC 136, and NGC 7245}}.
\newblock \bibinfo{journal}{Astronomy Letters} \bibinfo{volume}{36},
  \bibinfo{pages}{14--26}.
\newblock \DOIprefix\doi{10.1134/S1063773710010032}.
\bibitem[{{Haroon} et~al.(2014){Haroon}, {Ismail} and
  {Alnagahy}}]{2014Ap&SS.352..665H}
\bibinfo{author}{{Haroon}, A.A.}, \bibinfo{author}{{Ismail}, H.A.},
  \bibinfo{author}{{Alnagahy}, F.Y.}, \bibinfo{year}{2014}.
\newblock \bibinfo{title}{{Two MASS photometry of open star clusters: King 13
  and Berkeley 53}}.
\newblock \bibinfo{journal}{\apss} \bibinfo{volume}{352},
  \bibinfo{pages}{665--671}.
\newblock \DOIprefix\doi{10.1007/s10509-014-1990-z}.
\bibitem[{{Hensy}(2018)}]{2018JAsGe...7..180H}
\bibinfo{author}{{Hensy}, Y.H.M.}, \bibinfo{year}{2018}.
\newblock \bibinfo{title}{{Photometric and astrometric study of open cluster
  FSR 814 (Koposov 36) using SDSS/2MASS/PPMXL/Gaia DR2}}.
\newblock \bibinfo{journal}{\JAsGe} \bibinfo{volume}{7},
  \bibinfo{pages}{180--186}.
\newblock \DOIprefix\doi{10.1016/j.nrjag.2018.07.006}.
\bibitem[{{Janes} and {Phelps}(1994)}]{1994AJ....108.1773J}
\bibinfo{author}{{Janes}, K.A.}, \bibinfo{author}{{Phelps}, R.L.},
  \bibinfo{year}{1994}.
\newblock \bibinfo{title}{{The galactic system of old star clusters: The
  development of the galactic disk}}.
\newblock \bibinfo{journal}{\aj} \bibinfo{volume}{108},
  \bibinfo{pages}{1773--1785}.
\newblock \DOIprefix\doi{10.1086/117192}.
\bibitem[{{Jeffries} et~al.(2001){Jeffries}, {Thurston} and
  {Hambly}}]{2001A&A...375..863J}
\bibinfo{author}{{Jeffries}, R.D.}, \bibinfo{author}{{Thurston}, M.R.},
  \bibinfo{author}{{Hambly}, N.C.}, \bibinfo{year}{2001}.
\newblock \bibinfo{title}{{Photometry and membership for low mass stars in the
  young open cluster NGC 2516}}.
\newblock \bibinfo{journal}{\aap} \bibinfo{volume}{375},
  \bibinfo{pages}{863--889}.
\newblock \DOIprefix\doi{10.1051/0004-6361:20010918},
  \href{http://arxiv.org/abs/astro-ph/0107097}{{\tt arXiv:astro-ph/0107097}}.
\bibitem[{{King}(1962)}]{1962AJ.....67..471K}
\bibinfo{author}{{King}, I.}, \bibinfo{year}{1962}.
\newblock \bibinfo{title}{{The structure of star clusters. I. an empirical
  density law}}.
\newblock \bibinfo{journal}{\aj} \bibinfo{volume}{67}, \bibinfo{pages}{471}.
\newblock \DOIprefix\doi{10.1086/108756}.
\bibitem[{{Kumar} et~al.(2000){Kumar}, {Sagar}, {Rautela}, {Srivastava} and
  {Srivastava}}]{2000BASI...28..675K}
\bibinfo{author}{{Kumar}, B.}, \bibinfo{author}{{Sagar}, R.},
  \bibinfo{author}{{Rautela}, B.S.}, \bibinfo{author}{{Srivastava}, J.B.},
  \bibinfo{author}{{Srivastava}, R.K.}, \bibinfo{year}{2000}.
\newblock \bibinfo{title}{{Sky transparency over Naini Tal : A retrospective
  study}}.
\newblock \bibinfo{journal}{Bulletin of the Astronomical Society of India}
  \bibinfo{volume}{28}, \bibinfo{pages}{675--686}.
\bibitem[{{Landolt}(1992)}]{1922AJ....104..340L}
\bibinfo{author}{{Landolt}, A.U.}, \bibinfo{year}{1992}.
\newblock \bibinfo{title}{{UBVRI photometric standard stars in the magnitude
  range 11.5<V<16.0 around the celestial equator}}.
\newblock \bibinfo{journal}{\aj} \bibinfo{volume}{104}, \bibinfo{pages}{340}.
\bibitem[{{Lynga}(1982)}]{1982A&A...109..213L}
\bibinfo{author}{{Lynga}, G.}, \bibinfo{year}{1982}.
\newblock \bibinfo{title}{{Open clusters in our Galaxy}}.
\newblock \bibinfo{journal}{\aap} \bibinfo{volume}{109},
  \bibinfo{pages}{213--222}.
\bibitem[{{Marx} and {Lehmann}(1979)}]{1979AN....300..295M}
\bibinfo{author}{{Marx}, S.}, \bibinfo{author}{{Lehmann}, H.},
  \bibinfo{year}{1979}.
\newblock \bibinfo{title}{{Three color photometry of the open cluster AN King
  13}}.
\newblock \bibinfo{journal}{Astronomische Nachrichten} \bibinfo{volume}{300},
  \bibinfo{pages}{295--300}.
\newblock \DOIprefix\doi{10.1002/asna.19793000607}.
\bibitem[{{Mathis}(1990)}]{1990ARA&A..28...37M}
\bibinfo{author}{{Mathis}, J.S.}, \bibinfo{year}{1990}.
\newblock \bibinfo{title}{{Interstellar dust and extinction.}}
\newblock \bibinfo{journal}{\araa} \bibinfo{volume}{28},
  \bibinfo{pages}{37--70}.
\newblock \DOIprefix\doi{10.1146/annurev.aa.28.090190.000345}.
\bibitem[{{Pandey} et~al.(1997){Pandey}, {Durgapal}, {Bhatt}, {Mohan} and
  {Mahra}}]{1997A&AS..122..111P}
\bibinfo{author}{{Pandey}, A.K.}, \bibinfo{author}{{Durgapal}, A.K.},
  \bibinfo{author}{{Bhatt}, B.C.}, \bibinfo{author}{{Mohan}, V.},
  \bibinfo{author}{{Mahra}, H.S.}, \bibinfo{year}{1997}.
\newblock \bibinfo{title}{{Stellar contents of the open clusters Be 64 and Be
  69}}.
\newblock \bibinfo{journal}{\aaps} \bibinfo{volume}{122},
  \bibinfo{pages}{111--121}.
\newblock \DOIprefix\doi{10.1051/aas:1997296}.
\bibitem[{{Persson} et~al.(1998){Persson}, {Murphy}, {Krzeminski}, {Roth} and
  {Rieke}}]{1998AJ....116.2475P}
\bibinfo{author}{{Persson}, S.E.}, \bibinfo{author}{{Murphy}, D.C.},
  \bibinfo{author}{{Krzeminski}, W.}, \bibinfo{author}{{Roth}, M.},
  \bibinfo{author}{{Rieke}, M.J.}, \bibinfo{year}{1998}.
\newblock \bibinfo{title}{{A New System of Faint Near-Infrared Standard
  Stars}}.
\newblock \bibinfo{journal}{\aj} \bibinfo{volume}{116},
  \bibinfo{pages}{2475--2488}.
\newblock \DOIprefix\doi{10.1086/300607}.
\bibitem[{{Piskunov} et~al.(2006){Piskunov}, {Kharchenko}, {R{\"o}ser},
  {Schilbach} and {Scholz}}]{2006A&A...445..545P}
\bibinfo{author}{{Piskunov}, A.E.}, \bibinfo{author}{{Kharchenko}, N.V.},
  \bibinfo{author}{{R{\"o}ser}, S.}, \bibinfo{author}{{Schilbach}, E.},
  \bibinfo{author}{{Scholz}, R.D.}, \bibinfo{year}{2006}.
\newblock \bibinfo{title}{{Revisiting the population of Galactic open
  clusters}}.
\newblock \bibinfo{journal}{\aap} \bibinfo{volume}{445},
  \bibinfo{pages}{545--565}.
\newblock \DOIprefix\doi{10.1051/0004-6361:20053764},
  \href{http://arxiv.org/abs/astro-ph/0508575}{{\tt arXiv:astro-ph/0508575}}.
\bibitem[{{Rangwal} et~al.(2019){Rangwal}, {Yadav}, {Durgapal}, {Bisht} and
  {Nardiello}}]{2019MNRAS.490.1383R}
\bibinfo{author}{{Rangwal}, G.}, \bibinfo{author}{{Yadav}, R.K.S.},
  \bibinfo{author}{{Durgapal}, A.}, \bibinfo{author}{{Bisht}, D.},
  \bibinfo{author}{{Nardiello}, D.}, \bibinfo{year}{2019}.
\newblock \bibinfo{title}{{Astrometric and photometric study of NGC 6067, NGC
  2506, and IC 4651 open clusters based on wide-field ground and Gaia DR2
  data}}.
\newblock \bibinfo{journal}{\mnras} \bibinfo{volume}{490},
  \bibinfo{pages}{1383--1396}.
\newblock \DOIprefix\doi{10.1093/mnras/stz2642},
  \href{http://arxiv.org/abs/1909.08810}{{\tt arXiv:1909.08810}}.
\bibitem[{{Riess} et~al.(2018){Riess}, {Casertano}, {Yuan}, {Macri},
  {Bucciarelli}, {Lattanzi}, {MacKenty}, {Bowers}, {Zheng}, {Filippenko},
  {Huang} and {Anderson}}]{2018ApJ...861..126R}
\bibinfo{author}{{Riess}, A.G.}, \bibinfo{author}{{Casertano}, S.},
  \bibinfo{author}{{Yuan}, W.}, \bibinfo{author}{{Macri}, L.},
  \bibinfo{author}{{Bucciarelli}, B.}, \bibinfo{author}{{Lattanzi}, M.G.},
  \bibinfo{author}{{MacKenty}, J.W.}, \bibinfo{author}{{Bowers}, J.B.},
  \bibinfo{author}{{Zheng}, W.}, \bibinfo{author}{{Filippenko}, A.V.},
  \bibinfo{author}{{Huang}, C.}, \bibinfo{author}{{Anderson}, R.I.},
  \bibinfo{year}{2018}.
\newblock \bibinfo{title}{{Milky Way Cepheid Standards for Measuring Cosmic
  Distances and Application to Gaia DR2: Implications for the Hubble
  Constant}}.
\newblock \bibinfo{journal}{\apj} \bibinfo{volume}{861}, \bibinfo{pages}{126}.
\newblock \DOIprefix\doi{10.3847/1538-4357/aac82e},
  \href{http://arxiv.org/abs/1804.10655}{{\tt arXiv:1804.10655}}.
\bibitem[{{Salpeter}(1955)}]{1955ApJ...121..161S}
\bibinfo{author}{{Salpeter}, E.E.}, \bibinfo{year}{1955}.
\newblock \bibinfo{title}{{The Luminosity Function and Stellar Evolution.}}
\newblock \bibinfo{journal}{\apj} \bibinfo{volume}{121}, \bibinfo{pages}{161}.
\newblock \DOIprefix\doi{10.1086/145971}.
\bibitem[{{Spitzer} and {Hart}(1971)}]{1971ApJ...164..399S}
\bibinfo{author}{{Spitzer}, Jr., L.}, \bibinfo{author}{{Hart}, M.H.},
  \bibinfo{year}{1971}.
\newblock \bibinfo{title}{{Random Gravitational Encounters and the Evolution of
  Spherical Systems. I. Method}}.
\newblock \bibinfo{journal}{\apj} \bibinfo{volume}{164}, \bibinfo{pages}{399}.
\newblock \DOIprefix\doi{10.1086/150855}.
\bibitem[{{Stetson}(1987)}]{1987PASP...99..191S}
\bibinfo{author}{{Stetson}, P.B.}, \bibinfo{year}{1987}.
\newblock \bibinfo{title}{{DAOPHOT - A computer program for crowded-field
  stellar photometry}}.
\newblock \bibinfo{journal}{\pasp} \bibinfo{volume}{99},
  \bibinfo{pages}{191--222}.
\newblock \DOIprefix\doi{10.1086/131977}.
\bibitem[{{Subramaniam} and {Bhatt}(2007)}]{2007MNRAS.377..829S}
\bibinfo{author}{{Subramaniam}, A.}, \bibinfo{author}{{Bhatt}, B.C.},
  \bibinfo{year}{2007}.
\newblock \bibinfo{title}{{Photometric study of distant open clusters in the
  second quadrant: NGC 7245, King 9, King 13 and IC 166}}.
\newblock \bibinfo{journal}{\mnras} \bibinfo{volume}{377},
  \bibinfo{pages}{829--834}.
\newblock \DOIprefix\doi{10.1111/j.1365-2966.2007.11648.x},
  \href{http://arxiv.org/abs/astro-ph/0703075}{{\tt arXiv:astro-ph/0703075}}.
\bibitem[{{Yadav} et~al.(2008){Yadav}, {Kumar}, {Subramaniam}, {Sagar} and
  {Mathew}}]{2008MNRAS.390..985Y}
\bibinfo{author}{{Yadav}, R.K.S.}, \bibinfo{author}{{Kumar}, B.},
  \bibinfo{author}{{Subramaniam}, A.}, \bibinfo{author}{{Sagar}, R.},
  \bibinfo{author}{{Mathew}, B.}, \bibinfo{year}{2008}.
\newblock \bibinfo{title}{{Optical and near-infrared photometric study of the
  open cluster NGC 637 and 957}}.
\newblock \bibinfo{journal}{\mnras} \bibinfo{volume}{390},
  \bibinfo{pages}{985--996}.
\newblock \DOIprefix\doi{10.1111/j.1365-2966.2008.13740.x},
  \href{http://arxiv.org/abs/0810.1409}{{\tt arXiv:0810.1409}}.

\end{thebibliography}







\clearpage

\clearpage

\clearpage

\clearpage

\clearpage

\clearpage

\clearpage

\clearpage

\clearpage

\clearpage

\clearpage

\clearpage

\clearpage

\clearpage

\clearpage

\begin{table}
\centering
\caption{Log of observations for the cluster under study. King 13 was observed on 2$^{nd}$ December 2014.}
\vspace{0.5cm}
\begin{tabular}{cl}
\hline
Pass band  &Exposure Time \\
&(in seconds)\\
\hline
$V$&240$\times$2, 120$\times$2\\
$I$&120$\times$2, 60$\times$2\\
\hline
\end{tabular}
\label{tab1}
\end{table}

\clearpage

\begin{table}
\centering
\caption{Derived Standardization coefficients and its errors. $C$ and $Z$
are color coefficients and zeropoints respectively.}
\vspace{0.5cm}
\begin{tabular}{ccc}
\hline
Filter  &   $C$ & $Z$ \\
\hline

$V$&$-0.09\pm$0.004&$4.84\pm$0.008\\
$I$&$-0.11\pm$0.008&$5.35\pm$0.009\\
\hline
\end{tabular}
\label{tab2}
\end{table}

\clearpage

\begin{table}
\centering
\caption{The rms global photometric errors as a function of $V$ magnitude.}
\vspace{0.5cm}
\begin{tabular}{ccc}
\hline
$V$&$\sigma_{V}$&$\sigma_{I}$ \\
\hline
$14-15$&$0.04$&$0.05$ \\
$15-16$&$0.04$&$0.06$ \\
$16-17$&$0.05$&$0.08$ \\
$17-18$&$0.06$&$0.09$ \\
$18-19$&$0.06$&$0.10$ \\
$19-20$&$0.07$&$0.12$ \\
\hline
\end{tabular}
\label{tab3}
\end{table}

\clearpage

\begin{table}
\centering
\caption{Differences in $V$ and $(V-I)$ between \citet{2010AstL...36...14G}  and our study. The
standard deviation for the difference in each magnitude bin is also given in the parentheses.}
\vspace{0.5cm}
\begin{tabular}{crr}
\hline
$V$&$\Delta{V}$&$\Delta({V-I})$ \\
\hline
$13-14$&$-0.01~ (0.01)$&$0.02~ (0.02)$ \\
$14-15$&$-0.02~ (0.02)$&$-0.02~ (0.04)$ \\
$15-16$&$-0.04~ (0.04)$&$-0.03~ (0.05)$ \\
$16-17$&$-0.05~ (0.05)$&$-0.05~ (0.07)$ \\
$17-18$&$-0.07~ (0.07)$&$ 0.06~ (0.10)$ \\
$18-19$&$-0.08~ (0.08)$&$ 0.07~ (0.12)$ \\
$19-20$&$-0.10~ (0.10)$&$ 0.09~ (0.13)$ \\
\hline
\end{tabular}
\label{match_error}
\end{table}

\clearpage

\begin{table}
\centering
\caption{Structural parameters of the cluster King 13. Background and central density are in the unit
of stars per arcmin$^{2}$. $r_c$ is in arcmin while $R_t$ is in pc.
}
\vspace{0.5cm}
\begin{center}
\small
\begin{tabular}{cccccc}
\hline
Name & $f_{0}$ &$f_{b}$& $r_{c}$& $R_{t}$ & $\delta_{c} $ \\
\hline
King 13 & $36.3$&$16.04$&$0.6$&$8.5$&$2.8$ \\
\hline
\end{tabular}
\label{tab4}
\end{center}
\end{table}

\clearpage

\begin{table}
\centering
\caption{The photometric completeness of the data in each
magnitude bin for the cluster King 13.
}
\vspace{0.5cm}
\begin{center} 
\small
\begin{tabular}{cccccc}
\hline
   V (mag)    &        Completeness    \\
\hline
14 - 15 &        0.99   \\\\
15 - 16 &        0.97   \\\\
16 - 17 &        0.95   \\\\
17 - 18 &        0.91   \\\\
18 - 19 &        0.75   \\\\
\hline
\end{tabular}
\label{comp}
\end{center}
\end{table}

\clearpage

\begin{table*}
\centering
\caption{ Various fundamental parameters of the cluster King 13.}
\vspace{0.5cm}
\begin{center}
\small
\begin{tabular}{cc}
\hline
Parameter &  King 13 \\
\hline
\\

Radius                  &  $3.2$  arcmin                         \\
Right Ascension         &  $2.54\pm0.01$ deg          \\
Declination             &  $61.19\pm0.01$ deg         \\
$\mu_{\alpha}cos\delta$ & $-2.8\pm0.2$ mas $yr^{-1}$\\
$\mu_{\delta}$     &  $-0.9\pm0.1$ mas $yr^{-1}$    \\
Age                &  $510\pm60$ Myr                  \\
$[Fe/H]$             &  $-0.03\pm0.01$ dex              \\ 
Metal abundance    &  $0.012$                          \\
$E(G_{BP}-G_{RP})$  &  $0.71\pm0.26$                   \\
E(B-V)             &  $0.80\pm0.2$ mag               \\
$A_{V}$            &  2.43                            \\
$R_{V}$            &  3.04                             \\
Distance modulus   &  $14.80\pm0.20$ mag               \\
Distance (From Isochrone fitting) &  $3.75\pm0.40$ Kpc               \\
Distance (From mean Parallax) &  $3.84\pm0.15$ Kpc               \\
$X_{\odot}$        &  3.68 kpc                        \\
$Y_{\odot}$        &  9.57 kpc                       \\
$Z_{\odot}$        &  -0.08 kpc                        \\
$R_{GC}$           &   11.23 kpc                       \\
Total Luminosity   &   $\sim 4.2$ mag                  \\
Cluster members    &   172                            \\
IMF slope          &   $1.46\pm0.31$                  \\
Total mass         &   $\sim 270 M_{\odot}$           \\
Average mass       &   $1.56 M_{\odot}$               \\
Relaxation time    &   7.5 Myr                        \\
\hline
\end{tabular}
\label{tab5}
\end{center}
\end{table*}

\clearpage
\end{document}